\documentclass[sigconf]{acmart} 

\usepackage{graphicx}
\usepackage{subfigure}
\usepackage{amsmath}
\usepackage{amsfonts}
\usepackage{amssymb}
\usepackage{latexsym}
\usepackage{moreverb}
\usepackage{booktabs}
\usepackage{indentfirst}
\usepackage{picins}
\usepackage[linesnumbered,ruled]{algorithm2e}
\usepackage{framed}
\graphicspath{{figures/}}

\newcommand{\Eq}[1]{Eq.~(\ref{#1})}

\newcommand{\Fig}[1]{Fig.~\ref{#1}}

\newcommand{\Table}[1]{Table~\ref{#1}}

\newcommand{\Sec}[1]{Section~\ref{#1}}
\newcommand{\Tab}[1]{Table~\ref{#1}}

\newcommand{\Algo}[1]{Algorithm~\ref{#1}}

\acmYear{2018}
\copyrightyear{2018}
\setcopyright{acmcopyright}
\acmConference{MobiSys'18}{June 10--15, 2018}{Munich, Germany}
\acmDOI{978-1-4503-5720-3/18/06}
\acmISBN{https://doi.org/10.1145/3210240.3210320}

\begin{document}

\title[CrowdEstimator]{CrowdEstimator: Approximating Crowd Sizes with Multi-modal Data for Internet-of-Things Services}


\author{Fang-Jing~Wu}
\affiliation{
  \institution{TU Dortmund University, Dortmund, Germany}
}
\email{fang-jing.wu@tu-dortmund.de}

\author{G\"urkan~Solmaz}
\affiliation{
  \institution{NEC Laboratories Europe, Heidelberg, Germany}
}
\email{gurkan.solmaz@neclab.eu}

\begin{abstract}
Crowd mobility has been paid attention for the Internet-of-things (IoT) applications. This paper addresses the \emph{crowd estimation problem} and builds an IoT service to share the crowd estimation results across different systems. The crowd estimation problem is to approximate the crowd size in a targeted area using the observed information (e.g., Wi-Fi data). This paper exploits Wi-Fi probe request packets (``Wi-Fi probes" for short) broadcasted by mobile devices to solve this problem. However, using only Wi-Fi probes to estimate the crowd size may result in inaccurate results due to various environmental uncertainties which may lead to \emph{crowd overestimation or underestimation}. Moreover, the ground-truth is unavailable because the coverage of Wi-Fi signals is time-varying and invisible. This paper introduces auxiliary sensors, \emph{stereoscopic cameras}, to collect the near ground-truth at a specified calibration choke point. Two calibration algorithms are proposed to solve the crowd estimation problem. The key idea is to calibrate the Wi-Fi-only crowd estimation based on the correlations between the two types of data modalities. Then, to share the calibrated results across systems required by different stakeholders, our system is integrated with the FIWARE-based IoT platform. To verify the proposed system, we have launched an indoor pilot study in the Wellington Railway Station and an outdoor pilot study in the Christchurch Re:START Mall in New Zealand. The large-scale pilot studies show that stereoscopic cameras can reach minimum accuracy of $85\%$ and high precision detection for providing the near ground-truth. The proposed calibration algorithms reduce estimation errors by $43.68\%$ on average compared to the Wi-Fi-only approach.
\end{abstract}



\maketitle

\footnotesize \copyright ACM 2018. This is the author's version of the work. It is posted here for your personal use. Not for redistribution. The definitive Version of Record was published in ACM MobiSys'18, http://doi.org/10.1145/3210240.3210320

\section{Introduction}
Internet-of-Things (IoT) enrich confluence of communication technologies, cyber-physical systems, and data analytics have boosted many promising applications such as health-care systems~\cite{Xiong2017_iBlink}, indoor tracking~\cite{Mao2017_Drone}, urban mobility monitoring~\cite{Wu2017_ReachableRegions}, and social inference~\cite{Solmaz2017-GroupMobility}. As the human mobility becomes an important aspect in many smart-city applications, data from mobile devices has been paid much attention. There exist two mechanisms for collecting mobility data from mobile devices: \emph{opt-in data contribution} using mobile applications~\cite{Al-Turjman16}~\cite{Kantarci14} and \emph{pervasive sniffing} by overhearing wireless packets broadcasted by mobile devices. However, the opt-in rate in the former mechanism affects the data quantity and quality, and it may suffer from lack of bootstrapping data in large-scale urban applications. Thus, this paper considers the pervasive sniffing, where multiple \emph{Wi-Fi sniffers} are deployed in a targeted area to capture Wi-Fi packets from nearby mobile devices. The interested type of Wi-Fi packets in this paper is \emph{Wi-Fi probe request packets} (``Wi-Fi probes" for short) which are used to search for available Wi-Fi networks in the proximity. Since Wi-Fi probes indicate the appearance of mobile users, they provide clues for estimating the crowd size (i.e., number of people) in the area covered by the Wi-Fi sniffer.

This paper exploits Wi-Fi probes to solve \emph{the crowd estimation problem}. The crowd estimation problem is to approximate the crowd size in a targeted area using limited information (e.g., Wi-Fi probes, sensor readings, and videos). However, using \emph{only} Wi-Fi probes to count number of detected mobile devices for solving the crowd estimation problem may result in inaccurate results due to various uncertainties in mobility behaviours, physical environments (e.g., obstacles), radio interference, and dynamic intervals of captured Wi-Fi probes depending on mobile device usage and moving speeds. The Wi-Fi-only crowd estimation can further result in \emph{crowd underestimation} or \emph{crowd overestimation}. Furthermore, since ranges of Wi-Fi signals are invisible and time-varying, the ground-truth of crowd size is not available especially for large-scale deployment when \emph{only} Wi-Fi probes are considered.

Therefore, this paper considers not only Wi-Fi sniffers but also additional auxiliary sensors which are able to collect the \emph{near ground-truth} with very high accuracy at a specified calibration choke point (where most people are expected to pass through) for further calibrating the crowd estimation results using the multi-modal data sources. Even if the auxiliary sensors are not supposed to be $100\%$ accurate, they still provide close to actual results at the calibration choke point. Stereoscopic cameras are introduced as auxiliary sensors at the calibration choke point, where the computer vision-based people counting provides the near ground-truth. However, deploying many stereoscopic cameras in a large-scale area results in high-costs and may raise privacy concerns in certain regions. The proposed system includes only a few stereoscopic cameras which are deployed at the calibration choke point to compensate for Wi-Fi-only crowd estimation and perform calibration. The key idea is to learn the correlations between the Wi-Fi-only and the camera-based crowd estimation results at the calibration choke point and further apply the correlations to the neighboring areas monitored only by Wi-Fi sniffers without stereoscopic cameras. However, the correlations change over time due to unpredictable uncertainties in the environment and the human mobility behaviours. Thus, two adaptive crowd estimation algorithms are proposed to dynamically learn the correlations and perform calibration in real-time.

Two pilot studies are launched for multiple stakeholders including IoT application developers, end-users, governmental organizations (such as city councils), and enterprises. Therefore, we build the crowd estimation service using our ``in-house'' FIWARE-based IoT platform for exposing the estimated crowd sizes. The IoT platform provides real-time service endpoint to external systems using the light-weight \emph{IoT broker} with higher throughput, which we call \emph{Thin Broker}~\cite{FogFlow}. Multiple applications are developed by different stakeholders to access the crowd estimation results in our two pilots through the Thin Broker. So, the multi-modal crowd estimation results can be broadly and transparently shared across IoT systems.

The proposed system supports two pilot studies in an outdoor pedestrian shopping mall and in an indoor train station in New Zealand. The outdoor pilot study in the Re:START shopping mall in Christchurch indicates that it suffers from the crowd overestimation problem when Wi-Fi-only crowd estimation is considered. The indoor pilot study in the Wellington Railway Station indicates that the crowd underestimation and overestimation problems appear alternatively during weekdays and during weekends when Wi-Fi-only crowd estimation technology is applied. Based on the measurements in large-scale pilot studies, the people counting by the stereoscopic cameras can reach minimum accuracy of $85\%$, and the high precision detection results can serve as the near ground-truth. Based on the correlations between Wi-Fi-only crowd estimation and people counting by the stereoscopic cameras at the calibration choke point, the proposed multi-modal calibration algorithms can reach a maximum normalized root mean square error of 0.25 and can significantly reduce estimation errors by $43.68\%$ on average compared to the Wi-Fi-only approach. Also, the calibrated results align with the daily and weekly patterns in the near ground-truth.

\section{Related Work}\label{Sec:RelatedWork}
Crowd mobility analytics focuses on understanding people distributions and their movements in targeted areas. \emph{Computer vision-based approaches}, radio-based approaches, and \emph{Wi-Fi-based approaches} are considered to address this issue.

Computer vision-based approaches \cite{Fu2015_crowdDensityNeuralNetworks}\cite{Li2014_CrowdedSceneSurvey}\cite{ncs_CrowdDetection2}\cite{Idrees_DenseCrowds} perform classification based on the features learned from images or videos to detect people. In \cite{Fu2015_crowdDensityNeuralNetworks}, neural networks are considered to estimate the density of crowds for improving the detection accuracy and speeds. Crowd detection technology using video content analytics is used in \cite{ncs_CrowdDetection2}. The work in \cite{Idrees_DenseCrowds} focuses on detecting dense crowds in images. While various computer vision-based approaches are proposed \cite{Li2014_CrowdedSceneSurvey}, these approaches may compromise personal privacy. Furthermore, the usage of cameras is restricted in different countries depending on the regulations and the nature of the locations where the cameras would be installed.

Radio-based approaches \cite{Xu2013_RSSI}\cite{Depatla2015_RSSI}\cite{Depatla2018_RSSI}\cite{Xi2014_CIS} exploit natures of signal propagation such as received signal strength indicators (RSSIs), channel statuses, and multi-paths to estimate the crowd size in a small area. RSSIs are taken into account in \cite{Xu2013_RSSI} for counting people and localization in an indoor office. The work in \cite{Depatla2015_RSSI}, people counting are conducted based on RSSIs between a static pair of transmitter and receiver along a line-of-sight path. In \cite{Depatla2018_RSSI}, the transmitter and receiver are deployed behind the walls, and the RSSIs are used to estimate number of people in the area in-between.  Recently, compared to RSSIs, channel state information (CSI) is more sensitive to moving objects and is used for counting people in \cite{Xi2014_CIS}. However, the above approaches target a small-scale environment.

Wi-Fi-based approaches \cite{Chilipirea2016_MonitoringCrowds}\cite{Acery2016_CrowdBehavior_WiFi}\cite{Mashhadi_2016_SpaceSyntax}\cite{Li2015_SenseFlow}\cite{Weppner2016_CrowdCondition} by analyzing wireless packets provide more flexible and low-cost options to perform crowd detection and human mobility monitoring in large-scale environments. The work in \cite{Chilipirea2016_MonitoringCrowds} proposes some filtering algorithms to handle uncertain and noisy data from Wi-Fi sniffers due to MAC address randomization, overlapping coverage between Wi-Fi sniffers, and high variance in Wi-Fi sensing ranges. In \cite{Acery2016_CrowdBehavior_WiFi}, Wi-Fi sniffers are deployed in an industrial exhibition to capture the Wi-Fi probes from mobile devices of attendees, and mobility patterns in each monitored zone are analyzed such as the number of unique MAC addresses, the number of Wi-Fi probes, and the brand statistics of mobile devices in each zone. The work in \cite{Mashhadi_2016_SpaceSyntax} extends \cite{Acery2016_CrowdBehavior_WiFi} to analyze not only crowd dynamics but also correlations between the spatial configuration and entrepreneurial opportunities in these zones based on attendees' mobility. In \cite{Li2015_SenseFlow}, the people flows are analyzed using Wi-Fi probes based on a sequence of frequently visited sensing zones. The work in \cite{Weppner2016_CrowdCondition} analyzes the RSSIs in the captured Wi-Fi probes to perform localization and further estimates crowd density in the monitored areas. Wi-Fi sniffing technology has been taken into account in some industrial products for crowd detection and monitoring \cite{SensorInsight}\cite{IPI_CrowdMonitoring}\cite{ncs_CrowdDetection1}.

There exist key technical differences of our proposed approach compared to the ones aforementioned studies. First, the crowd estimation problem using only Wi-Fi probes is raised though large-scale experimental observations and real-world pilots, where the crowd overestimation and underestimation are investigated. Second, the key technical breakthrough is to combine both Wi-Fi-based and computer vision-based approaches for addressing the crowd estimation problem so that they can compensate the essential limitations of each other. The Wi-Fi-based approach can support large-scale observations with lower-cost deployment efforts and less restriction in privacy, whereas the computer vision-based approaches can provide higher accuracy observations in small-scale areas. Third, since the proposed algorithms are lightweight without a prior learning phase, it can adapt to dynamic changes of crowds mobility in real-time and large-scale applications. Finally, the proposed system is integrated with the IoT platform to provide a reliable and real-time service for various stakeholders across IoT systems.

\section{The Crowd Estimation Problem}\label{Sec:ProblemStatement}
\subsection{Preliminary: Wi-Fi Service Discovery}
A mobile device can operate in either the \emph{passive scanning} mode or the \emph{active scanning} mode, as defined in \cite{IEEE802.11Standard}, to connect to Wi-Fi networks. When a mobile device operates in the passive scanning mode, it listens beacons from access points to connect to a Wi-Fi network. On the contrary, when a mobile device operates in the active scanning mode, it broadcasts Wi-Fi probes to look for available Wi-Fi networks. Compared to the passive scanning mode, mobile devices operating in the active scanning mode take shorter time to connect to Wi-Fi networks. Therefore, the active scanning mode has been implemented in most of mobile devices. Each Wi-Fi probe includes the source address which is the MAC address of the mobile device, the destination address, and a sequence number in its management frame. Nowadays, off-the-shelf mobile devices broadcast Wi-Fi probes depending on the usage of mobile device for reducing energy consumption. The intervals of Wi-Fi probes range from a few seconds to 120 seconds. For example, when a mobile device is in a sleep mode, the intervals are longer.

\begin{figure}
\centering
\includegraphics[width=\columnwidth]{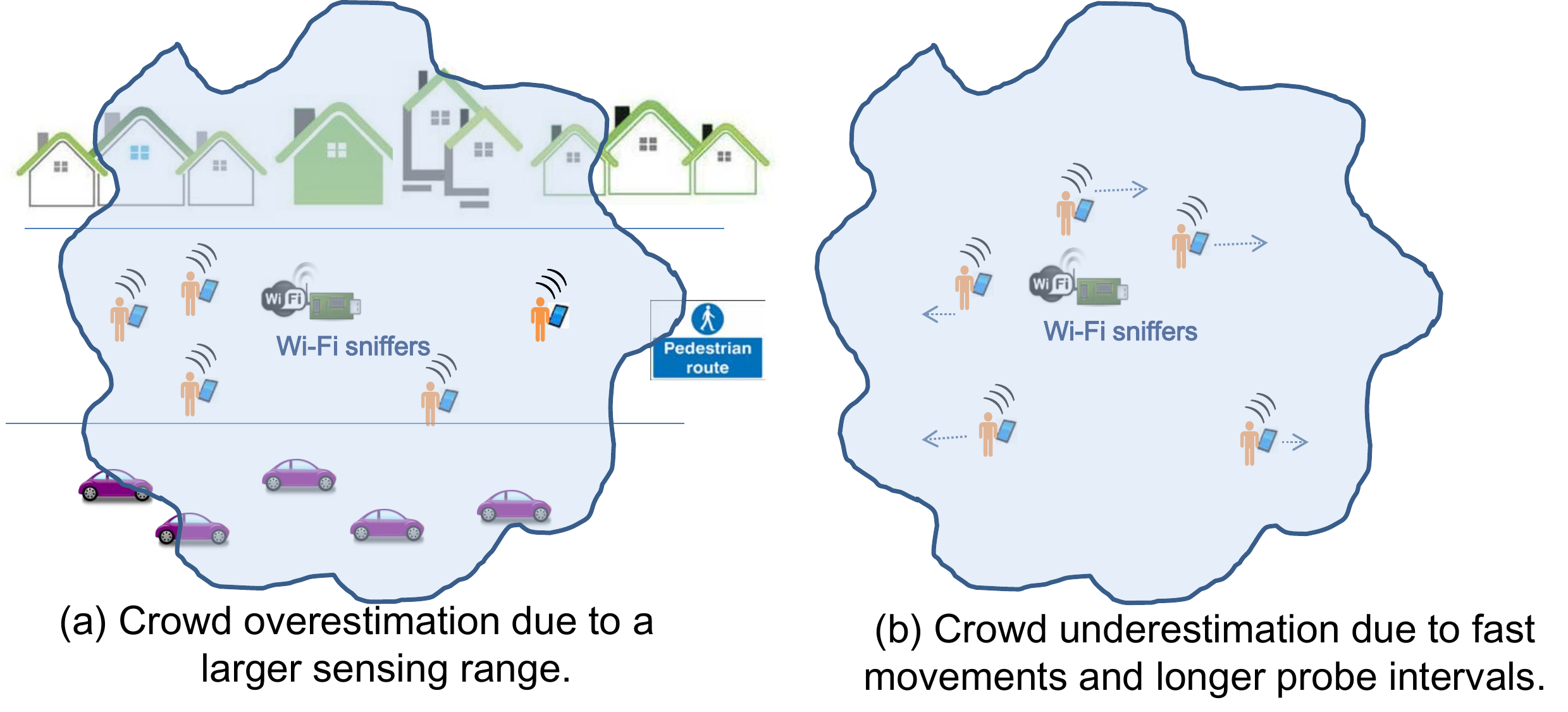}
\caption{The crowd overestimation and underestimation scenarios.} \label{Fig:Overestimation_and_underestimation}
\end{figure}

\subsection{Challenges of Crowd Estimation}
This paper uses \emph{Wi-Fi sniffing technology} to estimate the crowd size in targeted areas. A Wi-Fi sniffer is capable of overhearing all types of Wi-Fi packets and picking up Wi-Fi probes. Wi-Fi probes indicate the appearance of mobile devices even though the mobile devices do not connect to Wi-Fi networks. Intuitively, for a given area monitored by a Wi-Fi sniffer, the crowd size in the area can be estimated by counting the number of unique MAC addresses. However, the accuracy cannot be guaranteed due to the following reasons when \emph{only} Wi-Fi probes are taken into account.\\
\emph{$\bullet$ Unknown ground-truth}: Since the sensing ranges of Wi-Fi sniffers are invisible and they vary over time, the actual ground-truth is unknown. Nowadays, many of smart-city applications suffer from the similar issue especially for large-scale deployment. Furthermore, since it is hard to know who carries which mobile device with which MAC address, it is hard to verify the ground truth. Therefore, when the actual ground-truth is unknown, it is hard to evaluate the accuracy of the crowd estimation results based on only the total number of captured MAC addresses.\\
\emph{$\bullet$ Crowd overestimation} due to a larger sensing range: When the targeted area is much narrower than the Wi-Fi sniffing range (e.g., a pedestrian shopping area), the vehicle traffic or people passing by the neighborhood but not walking through the targeted area may be counted. Therefore, the estimated crowd size is more than the actual crowd size. \Fig{Fig:Overestimation_and_underestimation} (a) shows the crowd overestimation scenario using only the Wi-Fi sniffing technology.\\
\emph{$\bullet$ Crowd underestimation} due to longer probe intervals: Crowds make fast movements in some situations (e.g., bad weather conditions) or in some special environments (e.g. train stations). In such kind of situations, the probe intervals become longer because people normally do not check their mobile deivces when they are in a hurry. In this case, people have already moved out of the sensing range of the Wi-Fi sniffer before their mobile phones broadcast a Wi-Fi probe. As a result, the estimated crowd size based on the number of captured MAC addresses can be much less than the actual crowd size. \Fig{Fig:Overestimation_and_underestimation} (b) shows a scenario of the crowd underestimation problem when only the Wi-Fi sniffing technology is considered.

\section{Cross-modal Crowd Estimation}\label{Sec:Cross-modalCrowdEstimation}

\begin{figure}
\centering
\includegraphics[width=\columnwidth]{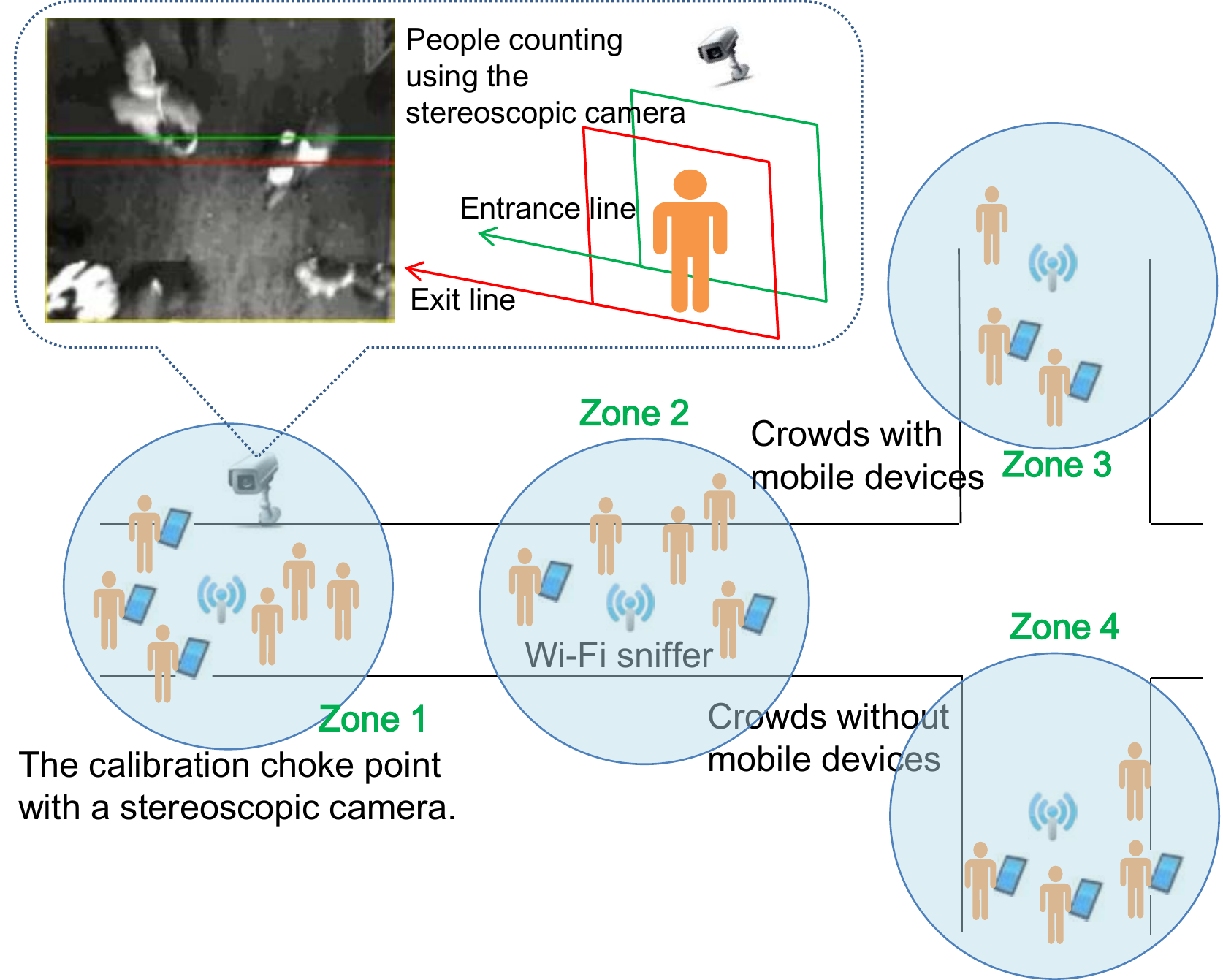}
\caption{An illustration of cross-modal crowd estimation.} \label{Fig:cross-modal}
\end{figure}

Assume that multiple Wi-Fi sniffers are deployed in an environment without many entrances such as a train station. Each Wi-Fi sniffer captures Wi-Fi probes in its sensing zone. We consider a \emph{calibration choke point}, where a Wi-Fi sniffer and a reliable stereoscopic camera are deployed for finding the correlations between the two data modalities. \Fig{Fig:cross-modal} illustrates the multi-modal deployment, where the calibration choke point is placed in Zone 1. The stereoscopic camera is mounted to cover the street so that people walking through can be captured and counted. An entrance line and an exit line are defined for determining the moving directions of the people using the computer vision-based technology. While the crowd estimation based on the information from the Wi-Fi sniffer is not precise, the stereoscopic camera is capable of providing precise number of people walking through the monitored area which can serve as the \emph{near ground-truth}. The near ground-truth using the stereoscopic camera can be verified by manually counting people in the captured videos. Thus, we can learn how reliable the stereoscopic camera's results are, and it can help to perform further calibration.

The key idea of the proposed cross-modal crowd estimation approach is to find the \emph{correlations} between the number of MAC addresses captured by the Wi-Fi sniffer and the number of people counted by the stereoscopic camera at the calibration choke point. Then, we apply the correlations to Wi-Fi-only crowd estimation results in other sensing zones without a stereoscopic camera. Since the sensing zones covered by these Wi-Fi sniffers are close to each other, the combinations of moving paths from one zone to the others are limited. For example, Zone 1 and Zone 2 are in the same main street, and the crowd distributions in the two zones are similar to each other in temporal and spatial domains. Therefore, the crowd distributions and correlations in those sensing zones are assumed the same because they are next to each other in the same proximity. The correlations can be applied to these sensing zones which do not have stereoscopic cameras, and it can further calibrate Wi-Fi-only results for all of the three zones based on the precise near ground-truth. Although the calibrated results for Zone 2 is more accurate compared to Zone 3 and Zone 4 which are located in the side roads, it still provides higher accuracy, as presented in \Sec{Sec:PilotStudies}, when these zones are closely located. Note that the Wi-Fi sniffers and stereoscopic cameras are used in this paper to compensate for each other's essential limitations. Wi-Fi sniffers provide a low-cost and privacy-preserving solution for large-scale monitoring, whereas stereoscopic cameras can provide more precise results in a small-scale area.

However, the correlations between the results from the two types of data sources vary over time and may change dynamically due to some uncertainties such as weather conditions, festivals, weekdays, and weekends. Therefore, re-learning the correlations to adapt to different environmental conditions becomes a challenge. This paper proposes \emph{adaptive} crowd estimation algorithms in \Sec{SubSection:Adaptive_Crowd_Estimation} to dynamically re-learn the correlations between two data modalities so that the correlations can be updated in real-time.

\subsection{System Architecture}

\begin{figure}
\centering
\includegraphics[width=\columnwidth]{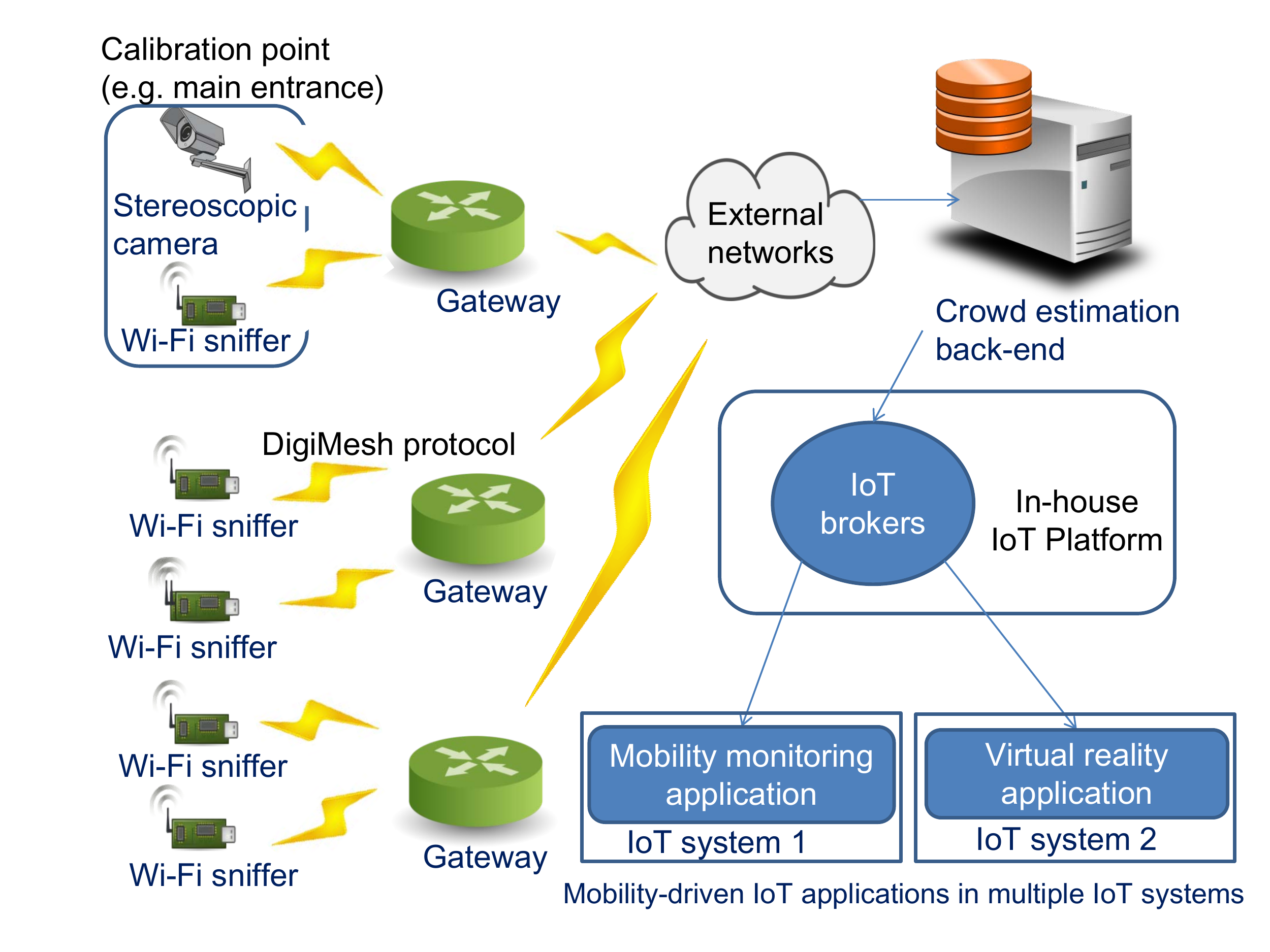}
\caption{The system architecture.} \label{Fig:SysArchi}
\end{figure}

\Fig{Fig:SysArchi} illustrates the system architecture of the proposed crowd estimation service.  Multiple Wi-Fi sniffers are deployed in the targeted environment, and a calibration choke point is selected based on human domain knowledge for deploying a stereoscopic camera. For example, the main entrance in a pedestrian shopping area or in a train station can serve as the calibration choke point because most of pedestrians appear there. The calibration choke point provides richer information to compute the correlations between detected events by the two different data sources (i.e., the Wi-Fi sniffer and the stereoscopic camera). These data sources are connected to their local gateways through the DigiMesh protocol. The local gateways report collected data to the crowd estimation back-end infrastructure through external networks (i.e., 3G networks). The proposed \emph{cross-modal crowd estimation module} is implemented in the back-end infrastructure. It publishes the calibrated crowd estimation results to the multiple instances of IoT broker deployed in the in-house IoT platform so that third-party IoT systems can query the IoT brokers for developing their own applications.

\subsection{Data Quality for the Near Ground-Truth}
\begin{figure}
\centering
\includegraphics[width=\columnwidth]{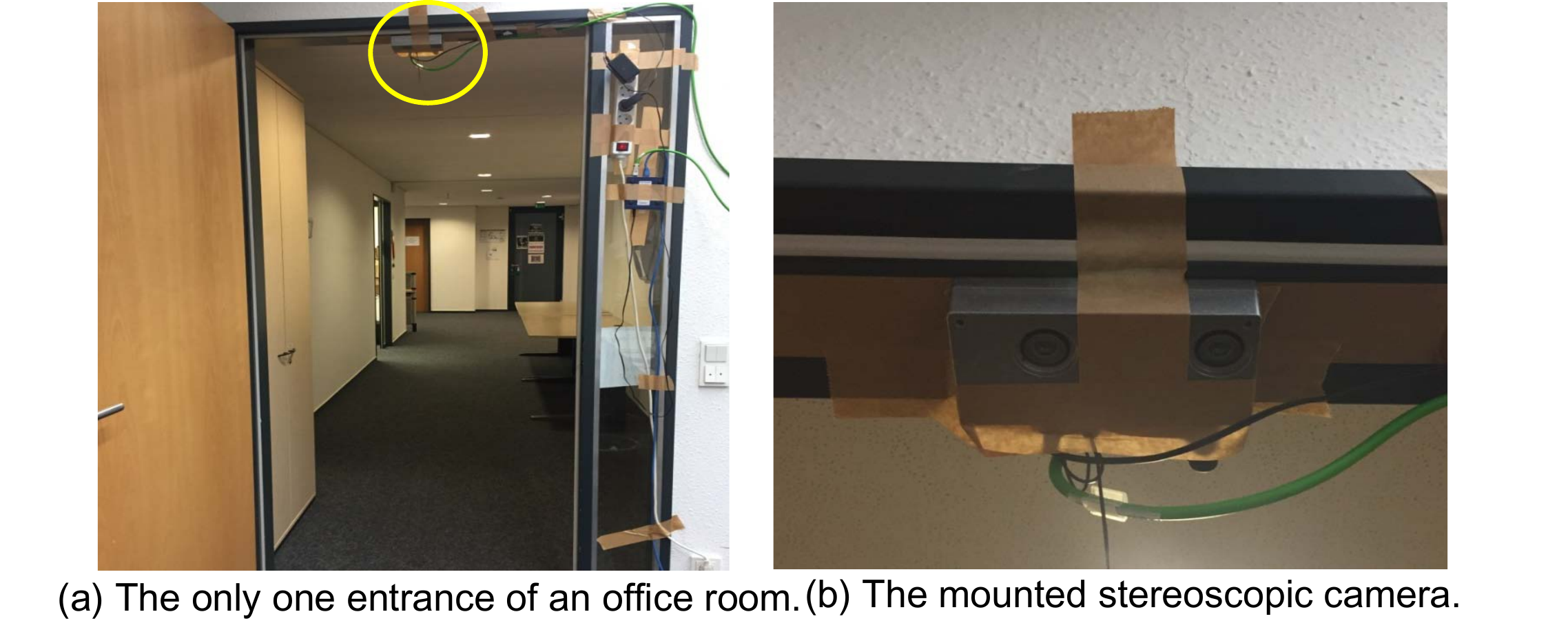}
\caption{Experimental setup for verification of the near ground-truth.} \label{Fig:lab_exp_camera}
\end{figure}

To evaluate the data quality for providing the near ground-truth, we conduct experiments in a lab environment to verify the \emph{accuracy} and the \emph{precision} of the stereoscopic camera with the computer vision-based software for people counting.
\begin{itemize}
 \item \emph{Accuracy} is the proximity of measurements to the true value. For example, if the actual number of people is 20, and the number of people detected by the stereoscopic camera is 3, then results provided by the stereoscopic camera is inaccurate.
  \item \emph{Precision} refers to the repeatability or reproducibility of the measurements. For example, considering the same example above, if the same experiment is repeated for 100 times, and the stereoscopic camera detects 3 people each time, then the results provided by the stereoscopic camera is very precise.
\end{itemize}
Based on these two metrics, we verify if the stereoscopic camera provides reliable and quality data to serve as the near ground-truth. \Fig{Fig:lab_exp_camera} shows our experimental setup in the lab environment. An office room with only one entrance is considered. An off-the-shelf stereoscopic camera is mounted on the entrance to monitor people moving in and from the office room.

First, to verify the accuracy of the stereoscopic camera, we consider two different scenarios: a static crowd and a dynamic crowd in the following two independent experiments. In the first experiment, 4 people are mostly static in the office room, but they still move in or out from time to time during the experiment. The total duration of this experiment is 3 hours. During the experiment, the 4 people make 63 movements (move-in or move-out events), and the 63 events are all detected by the the stereoscopic camera. However, two of these events are false detection because one person makes ambiguous movements (back and forth movements) under the stereoscopic camera. The accuracy in the first experiment with a static crowd is $96.8\%(=61/63)$. We conduct the second experiment with a dynamic crowds of 14 people. The 14 people walk in the monitored office room and then walk out of the office room after staying for a couple of minutes. The duration of the experiment is 20 minutes. During the experiment, people walk not exactly one by one, but sometimes 2 people walk together or two just after the other two because they talk to each other while walking. Moreover, they are allowed arbitrarily enter or leave the office room. In total, 41 events are detected in 20 minutes, and one of them is false detection. Therefore,the accuracy for the dynamic scenario is $97.5\%(=40/41)$.

Then, to verify the precision, we conducted the experiment, where a single person repeatedly moves in and out the office room for 10 times. The detection results of these repeated movements are all the same. Note that, compared to the stereoscopic camera, a Wi-Fi sniffer is not able to provide a good enough precision due to uncertain intervals of Wi-Fi probes. Specifically, although same movements are repeated by the same person, the detection results can be different depending on the intervals of Wi-Fi probes.

Based on the above observations, we assume that the stereoscopic camera can provide a baseline as the near ground-truth in bright environments. We further verify the accuracy in the large-scale pilot studies in \Sec{Sec:PilotStudies}.

\subsection{Adaptive Calibration Algorithms}\label{SubSection:Adaptive_Crowd_Estimation}

\begin{figure}
\centering
\includegraphics[width=\columnwidth]{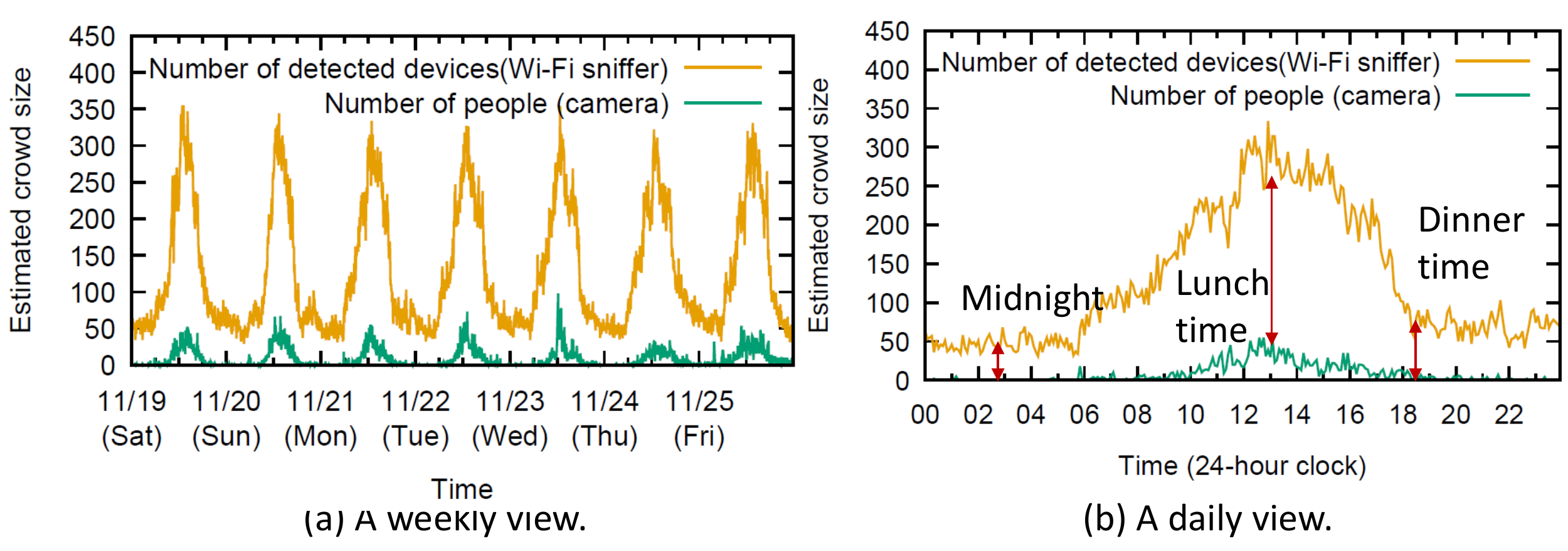}
\caption{Dynamic changes of correlations between the events detected by the two different data sources.} \label{Fig:Correlation-observations}
\end{figure}

The key idea of the proposed algorithm is to calibrate the Wi-Fi-only crowd estimation results using the correlations between events detected by the Wi-Fi sniffer and the stereoscopic camera at the calibration choke point. First, we conduct an experiment in the Re:START pedestrian shopping mall in Christchurch to make observations on correlations between events detected by the two types of data sources. A Wi-Fi sniffer and a stereoscopic camera are mounted at the main entrance gate to monitor pedestrians who enter and leave the pedestrian shopping mall. The duration of the experiment is 1 week. \Fig{Fig:Correlation-observations} shows the experimental results during the one week. As we can see in \Fig{Fig:Correlation-observations}~(a), the crowd estimation results using the different data modalities have similar patterns. However, the difference between the two results changes over time. \Fig{Fig:Correlation-observations}~(b) shows the daily pattern. The differences during the midnight, lunch time, and dinner time vary over time depending on uncertain and unknown environmental conditions.

Since the correlations dynamically changes, two algorithms: (1) \emph{dynamic proportional calibration} and (2) \emph{adaptive linear calibration}, are proposed to adaptively re-learn the correlations and perform the cross-modal calibration over time. The first algorithm proportionally moderates the crowd estimation results based on the ratios between camera's and Wi-Fi's detection results. The second algorithm modifies the typical linear regression to fit the latest training data points from the two types of detected events in the sense that the training data points are continuously updated over time. In the proposed algorithms, fixed-size time windows are used so that the correlations can be updated time window by time window. For a given time window $t_i$, let $C_0$ denote the set of move-in and move-out events detected by the stereoscopic camera during $t_i$, and let $W_k$, $k=0, \ldots, n$, denote the set of Wi-Fi probes captured by the Wi-Fi sniffer $k$ during $t_i$. Here, $W_0$ denotes the set of Wi-Fi probes captured by the Wi-Fi sniffer at the calibration choke point.

\subsubsection{Dynamic Proportional Calibration}
\Algo{Algo:ProportionalCalibration} presents the pseudocode of the dynamic proportional calibration algorithm. There are two phases in the proposed algorithm: (i) \emph{correlation update phase} and (ii) \emph{calibration phase}. First, in the correlation update phase, the number of mobile devices detected by the Wi-Fi sniffer at the calibration choke point during $t_i$ (denoted by $d_0$), is computed by \Algo{Algo:device-counting}. Then, the algorithm accumulates the total number of the move-in and move-out events detected by the stereoscopic camera during $t_i$, denoted by $y_0$. Here, move-in and move-out events are accumulated because pedestrians can come from two opposite directions. Therefore, the correlation coefficient $a_i$ during this time window can be calculated. Then, in the calibration phase, the correlation coefficient $a_i$ is applied to $d_1, ..., d_n$, which are the Wi-Fi-only crowd estimation results in the other sensing zones, to compute the corresponding calibrated results $d'_1, ..., d'_n$ respectively. \Algo{Algo:ProportionalCalibration} is executed every time window so that calibration can be adaptively performed based on the time-varying correlations.

\begin{algorithm}
\footnotesize
    \caption{Dynamic Proportional Calibration $(t_i, C_0, \{W_0,..., W_n\})$}
    \label{Algo:ProportionalCalibration}
    \SetKwInOut{Input}{Input}
    \SetKwInOut{Output}{Output}

    \Input{$C_0$ is the set of move-in and move-out events detected by the stereoscopic camera during $t_i$, and $\{W_0,..., W_n\}$ is Wi-Fi probes captured by Wi-Fi sniffers.}
    \Output{$d'_1, d'_2..., d'_n$.}

    \emph{//Correlation update phase}:\\
    $d_0$=Wi-Fi-based device counting $(W_0,t_i)$\;
    $y_0$= $e_{in}$+$e_{out}$, where $e_{in}$ is the total number of move-in events and $e_{out}$ is the total number of move-out events in $C_0$\;
    Compute the proportional function:  $y_0=a_i \cdot d_0$\;

    \emph{//Calibration phase}:\\
    \For{$W_k, k=1, 2, ... n$}{
        $d_k$=Wi-Fi-based device counting $(W_k,t_i)$\;
        $d'_k=a_i \cdot d_k$\;
    }
    return $d'_1, d'_2..., d'_n$\;
\end{algorithm}

\begin{algorithm}
\footnotesize
    \caption{Wi-Fi-based device counting $(W_k,t_i)$}
    \label{Algo:device-counting}
    \SetKwInOut{Input}{Input}
    \SetKwInOut{Output}{Output}
    \Input{$W_k$ which is the set of Wi-Fi probes captured by Wi-Fi sniffer $k$ during time window $t_i$.}
    \Output{$d_k$ which is the number of detected mobile devices during time window $t_i$.}
    $d_k$=0\;
    $D=\emptyset$\;
    \For{$p\in W_k$}{
        \If{MAC address indicated in p $\notin D$}
        {
            $d_k=d_k+1$\;
            $D=D \bigcup$ $\{$ MAC address indicated in p $\}$\;
        }
    }

    return $d_k$\;
\end{algorithm}

\subsubsection{Adaptive Linear Calibration}
The key idea of the adaptive linear calibration is to find a linear function that fits the set of given measurements captured by the Wi-Fi sniffer and stereoscopic cameras at the calibration choke point so that the linear function can be applied in other sensing zones to further calibrate the Wi-Fi-only crowd estimation results. However, the typical linear regression may lead to negative values (i.e., negative crowd sizes). In addition, the typical linear regression cannot adapt to the dynamic changes of correlations between Wi-Fi-only results and computer vision-based results at the calibration choke point. Therefore, an adaptive linear regression is designed as a modification of the typical linear regression. The key idea of the adaptive linear regression is to limit the number of training data points to the latest $q$ measurements. Based on the latest $q$ measurements, the linear least square method is applied to compute a linear function going through the origin. Let $(x_1, y_1), \ldots, (x_q, y_q)$ denote the measurements during the latest $q$ time windows at the calibration choke point. Here, $x_i$ is the number of mobile devices detected by the Wi-Fi sniffer during the time window $t_i$ at the calibration choke point, and $y_i$ is the total counts of people detected by the stereoscopic cameras during the time window $t_i$ at the calibration choke point. Assume that the correlations between Wi-Fi-only results and computer vision-based results at the calibration choke point follows the linear function $y=ax$ going through the origin. Therefore, the total sum of the vertical offsets from the linear function to those measurements can be computed by
\begin{align}\label{Eq:TotalDistance}
  L^2&=\sum_{i=1}^{q} (y_i-a\cdot x_i)^2. \notag  \\
     &=\sum_{i=1}^{q} y_i^2 -2a\sum_{i=1}^{q}x_i\cdot y_i +a^2 \sum_{i=1}^{q} x_i^2. \notag
\end{align}

Therefore, the linear function can be approximated by finding the minimum of $L^2$. The condition for $L^2$ to be a minimum is that

\begin{equation}
   \frac{\partial L^2}{\partial a}=0. \notag
\end{equation}
So, we have

\begin{equation}
   \frac{\partial L^2}{\partial a} = -2\sum_{i=1}^{q} x_i y_i + 2a \sum_{i=1}^{q} x_i^2=0. \notag
\end{equation}
Thus,
\begin{equation}\label{Eq:LinearCoefficient}
a=\frac{\sum_{i=1}^{q} x_i y_i}{\sum_{i=1}^{q} x_i^2}.
\end{equation}

\Algo{Algo:AdaptiveLinearCalibration} presents the adaptive linear calibration algorithm. The input of the algorithm contains the set of move-in and move-out events detected by the stereoscopic camera during the current time window $t_i$ (denoted by $C_0$), the Wi-Fi probes captured by these Wi-Fi sniffers (denoted by $\{W_0,..., W_n\}$) during the current time window $t_i$, a set of historical measurements (denoted by $O$), and the limited size of training data points (denoted by $q$). The set of historical measurements $O$ consists of the numbers of mobile devices detected by the Wi-Fi sniffer and the total numbers of people detected by the stereoscopic camera at the calibration choke point during the latest $q$ time windows $t_{i-1}..t_{i-q}$. Initially, $O=\varnothing$ and it is continuously updated when a new measurement arrives. The maximum size of $O$ is $q$. When the new events arrive, $d_0$ and $y_0$ are computed. Then, a new training data point $(d_0, y_0)$ is added into $O$. After $O$ is updated, if the size of $O$ is larger than the limited number of training data points $q$, the oldest training data point is removed from $O$. Then, the linear function is updated using \Eq{Eq:LinearCoefficient} based on the updated training data points $O$. Finally, for those sensing zones without stereoscopic cameras, this updated linear function is applied to the Wi-Fi-only crowd estimation results for calibration.

\begin{algorithm}
\footnotesize
    \caption{Adaptive Linear Calibration $(t_i, C_0, \{W_0,..., W_n\}, O=\{(x_1, y_1)\dots\}, q)$}
    \label{Algo:AdaptiveLinearCalibration}
    \SetKwInOut{Input}{Input}
    \SetKwInOut{Output}{Output}

    \Input{$C_0$ is the set of move-in and move-out events detected by the stereoscopic camera during $t_i$, and $\{W_0,..., W_n\}$ is Wi-Fi probe request packets captured by Wi-Fi sniffers.}
    \Output{$d'_1, d'_2..., d'_n$.}

    \emph{//Update training data points and linear function}:\\
    $d_0$=Wi-Fi-based device counting $(W_0,t_i)$\;
    $y_0$= $e_{in}$+$e_{out}$, where $e_{in}$ is the total number of move-in events and $e_{out}$ is the total number of move-out events in $C_0$\;

    Add $(d_0, y_0)$ into $O$ \\

    \If{$|O| < q$ }{
      exit\;
    }
    \ElseIf{$|O| > q$} {
        Remove the oldest measurement from $O$\;
    }

    Update the linear function: $y=a \cdot x$ based on \Eq{Eq:LinearCoefficient}\;

    \emph{//Calibration phase}:\\
    \For{$W_k, k=1, 2, ... n$}{
        $d_k$=Wi-Fi-based device counting $(W_k,t_i)$\;
        $d'_k=a\cdot d_k$\;
    }
    return $d'_1, d'_2..., d'_n$\;
\end{algorithm}


\begin{figure*}[h!]
\centering
\begin{tabular}{cc}
\includegraphics[width=0.5\linewidth]{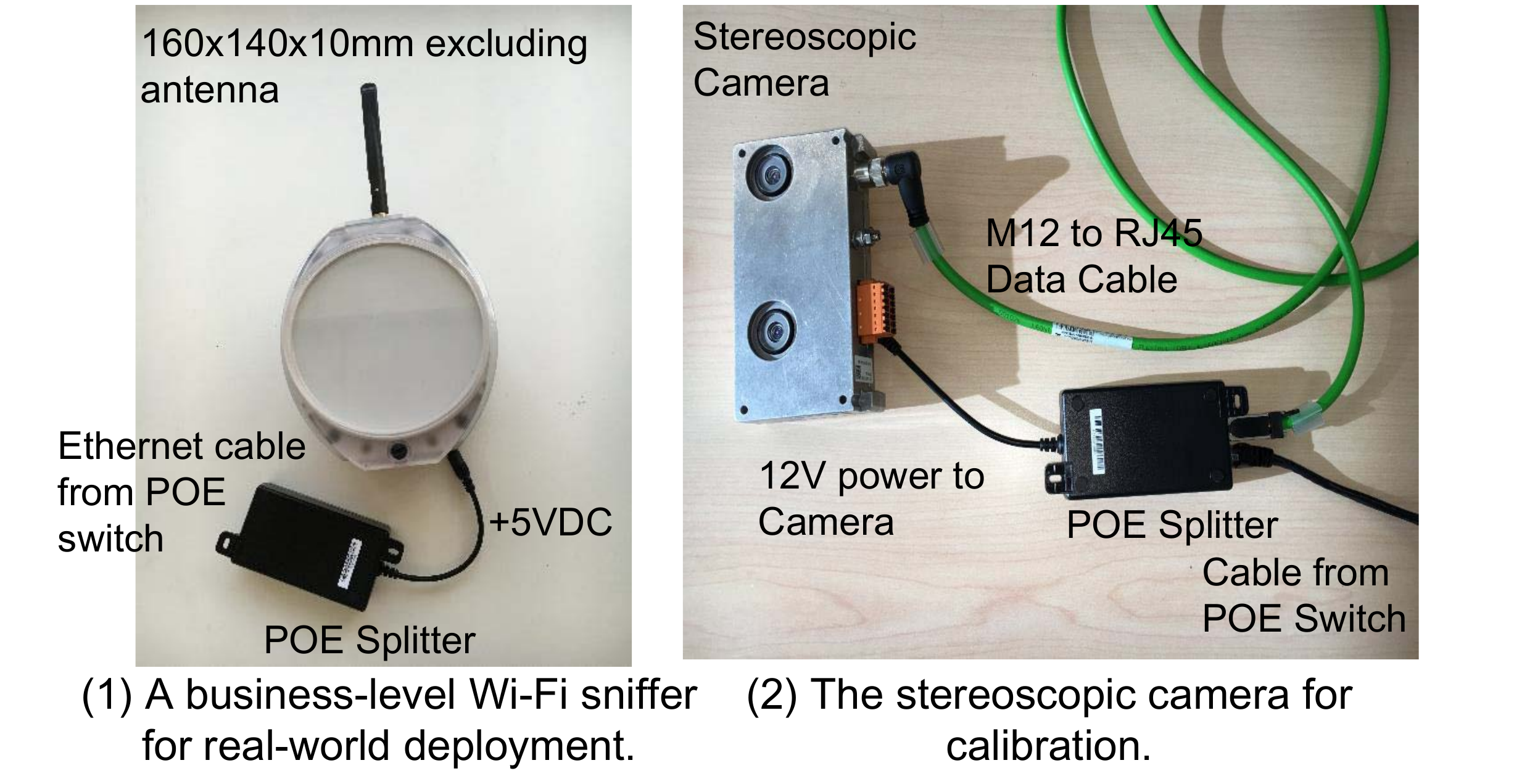} &
\includegraphics[width=0.47\linewidth]{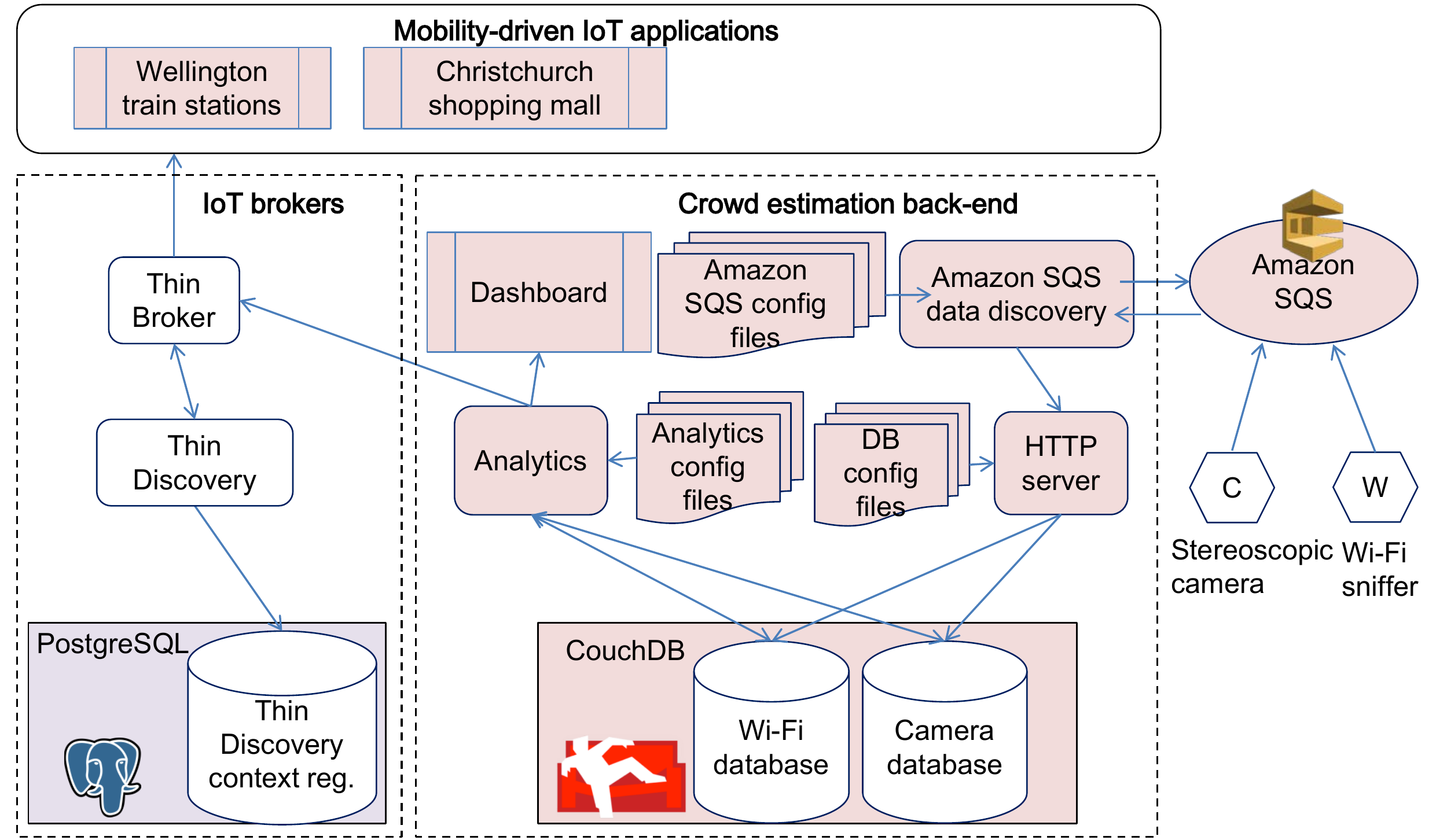} \\
(a) Hardware components. & (b) Software components.
\end{tabular}
\caption{Hardware and software components for both outdoor and indoor pilot studies.} \label{Fig:Hw-Sw}
\end{figure*}

\section{Pilot Studies}\label{Sec:PilotStudies}

\subsection{Hardware and Software Design}
We build business-level Wi-Fi sniffers to support our outdoor and indoor pilot studies. \Fig{Fig:Hw-Sw}(a)-(1) shows the developed Wi-Fi sniffer and \Fig{Fig:Hw-Sw}(a)-(2) shows the off-the-shelf stereoscopic camera~\cite{stereoscopic-camera}. \Fig{Fig:Hw-Sw}(b) shows the software components of the proposed service. The data sources (i.e., stereoscopic cameras and Wi-Fi sniffers) send sensing data to the Amazon Simple Queue Service (Amazon SQS) \cite{AWS_SQS}. Then, the Amazon SQS data discovery periodically makes queries to the Amazon SQS and posts to the HTTP server. The query interval of the Amazon SQS is specified in the Amazon SQS configuration files. To enable real-time service, the query interval for the pilot studies is 100 ms. After the HTTP server receives incoming real-time data, it immediately populates data into the Wi-Fi and camera databases, where a CouchDB is used in our implementation. The configurations of the databases for mapping attributes are specified in the database configuration files. Then, the data analytics component makes queries to both databases and performs cross-modal crowd estimation and calibration using the proposed adaptive algorithms in \Sec{SubSection:Adaptive_Crowd_Estimation}. The crowd estimation results are published to the IoT brokers so that multiple stakeholders involved in the pilot studies and their IoT applications can access the real-time results. The implemented IoT broker is called \emph{thin broker} ~\cite{FogFlow}, which efficiently handles the queries and subscriptions with with higher throughput. Meanwhile, the crowd estimation results are visualized in the dashboard. Two pilot studies in an outdoor pedestrian shopping mall in Christchurch and an indoor train station in Wellington are launched to verify the proposed cross-modal crowd estimation service.

\subsection{Integration with the IoT Platform}
Since the cross-modal crowd estimation results are required by multiple stakeholders (including city councils and industrial application developers) in the two pilot studies, the entire system is integrated with the in-house IoT platform which provides real-time service endpoint to external systems via the thin broker. Below, we describe the information model of crowd estimation results and then describe the components of the IoT platform that enables the cross-system crowd estimation service.

\subsubsection{NGSI-based Information Model}
The information model for crowd estimation is based on the FIWARE~\cite{FIWARE} {\em Next Generation Service Interfaces} (NGSI)~\cite{NGSI}. NGSI is a set standard interfaces for providing interoperability, information sharing, and system integration. NGSI context API ~\cite{NGSIMartin} enables access to a plethora of rich context information about users, places, events, and things. NGSI has become an open standard of FIWARE adopted by various smart cities all over the world. Therefore, the NGSI interface is used for crowd estimation service to achieve openness and interoperability between different applications, systems, and platforms. NGSI has an HTTP-based RESTful API which can have either JSON or XML formats for the message bodies. The NGSI-based information model is built on a structure which has \emph{entity}, \emph{attribute}, and \emph{metadata} relationships. Below, we specify the major entity: {\em Wi-Fi sniffer} and its attribute: {\em crowd estimation}. The first one represents a device, and the latter represents a data analytics result.

\begin{figure}
\centering
\includegraphics[width=0.7\linewidth]{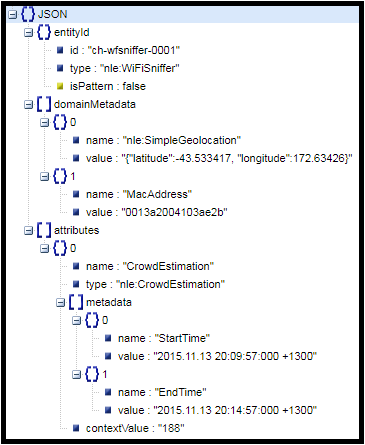}
\caption{An example of JSON format based on our information model.}
\label{Fig:JSON-CrowdEstimation}
\end{figure}

\Table{table:wifisniffertable} illustrates the data model for the Wi-Fi sniffer. Here, we specify the properties of the sniffer device. A Wi-Fi sniffer device is formally defined as the ``nle:WiFiSniffer''. The Wi-Fi sniffer has the ``nle:CrowdEstimation'' property as an attribute to represent the data analytics result. The {\em ``id''} and {\em ``type''} properties are modeled as entity id and type in the information model, whereas {\em ``nle:SimpleGeolocation''} and {\em ``MacAddress''} are modeled as the {\em domain metadata} of the entity. \Table{table:crowdestimationtable} defines the properties for the key attribute: crowd estimation. Crowd estimation attribute represents the data analytics results in the crowd estimation service. There are five basic properties of this attribute: {\em ``name''}, {\em ``type''}, {\em ``contextValue''}, {\em ``StartTime''}, and {\em ``EndTime''}. Context value represents the crowd estimation result for the given time window. The time window is specified by the start time and the end time. The {\em ``StartTime''} and {\em ``EndTime''} properties are modeled as the \emph{metadata of the attribute}. The defined NGSI-based information is converted to a JSON format for context exchanges across different IoT systems. \Fig{Fig:JSON-CrowdEstimation} shows an example JSON data which has the NGSI structure. 

\begin{table}
\scriptsize
\caption{The data model for the Wi-Fi sniffer entity.}
   \centering
        \begin{tabular} {|p{2.2cm}|p{1.0cm}|p{4.0cm}|}  \hline
      \textbf{Property}&\textbf{Expected type}&\textbf{Description} \\ \hline \hline
id & String & Entity's unique identifier\\
type & String & The type of entity. In this case, the defined value for the device is ``nle:WiFiSniffer''.\\
MacAddress & String & The MAC address of the Wi-Fi sniffer device.\\
nle:SimpleGeolocation	 & JSON & Contains location information (i.e., latitude and longitude) of the device.\\
nle:CrowdEstimation  & JSON & Attribute of the nle:WiFiSniffer where data analytical results reside.\\ \hline

	\end{tabular}

   \label{table:wifisniffertable}
   \end{table}

\begin{table}
\scriptsize
\caption{The properties of the crowd estimation attribute.}
   \centering
        \begin{tabular} {|p{1.7cm}|p{1.3cm}|p{4.0cm}|} \hline
\textbf{Property}&\textbf{Expected type}&\textbf{Description} \\ \hline \hline
name & String & The attribute's identifier. ``CrowdEstimation'' is used as the name.\\
type & String & The type of entity. In this case, the defined value is ``nle:CrowdEstimation''. \\
contextValue & Integer & The estimated crowd size. \\
StartTime & DateTime & The start time of the crowd estimation. \\
EndTime & DateTime &The end time of the crowd estimation. \\
\hline

	\end{tabular}
   \label{table:crowdestimationtable}
   \end{table}

\begin{figure*}[t!]
\centering
\begin{tabular}{cc}
\includegraphics[width=0.5\linewidth]{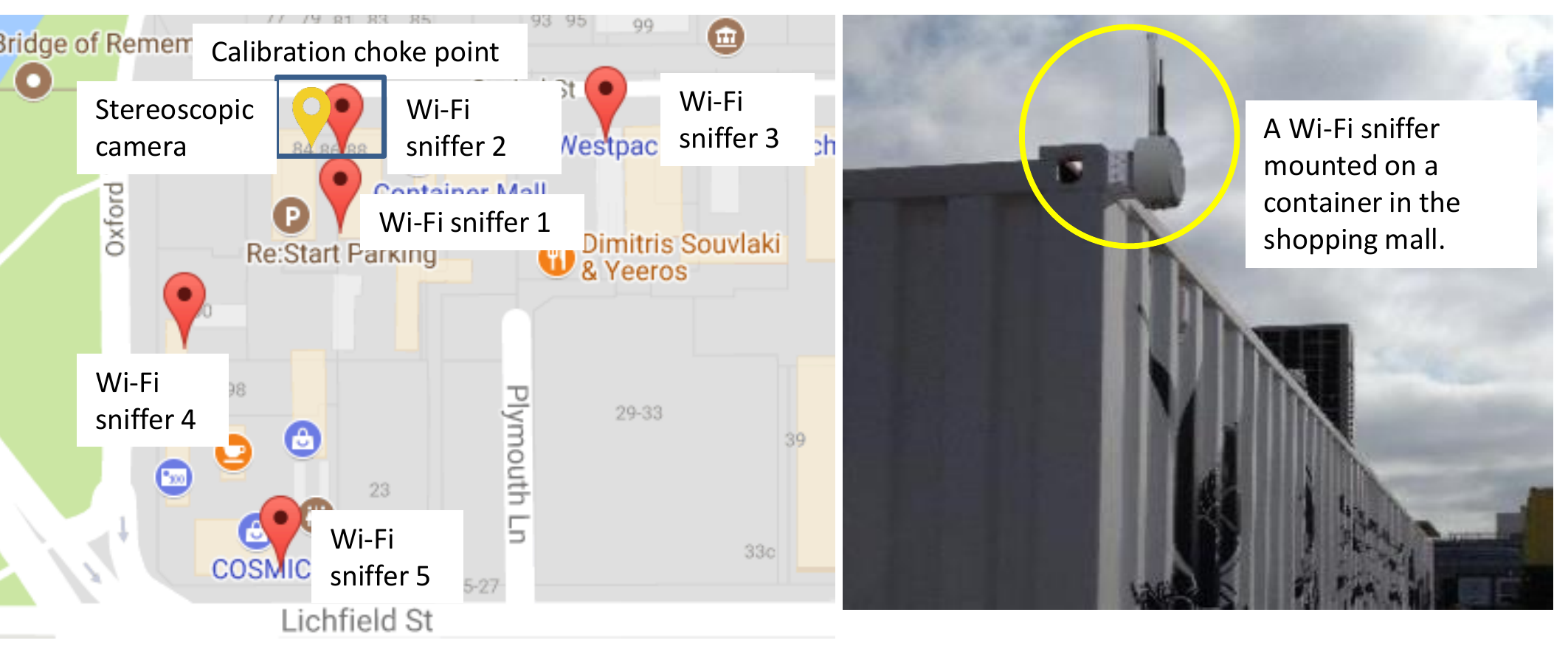} &
\includegraphics[width=0.4\linewidth]{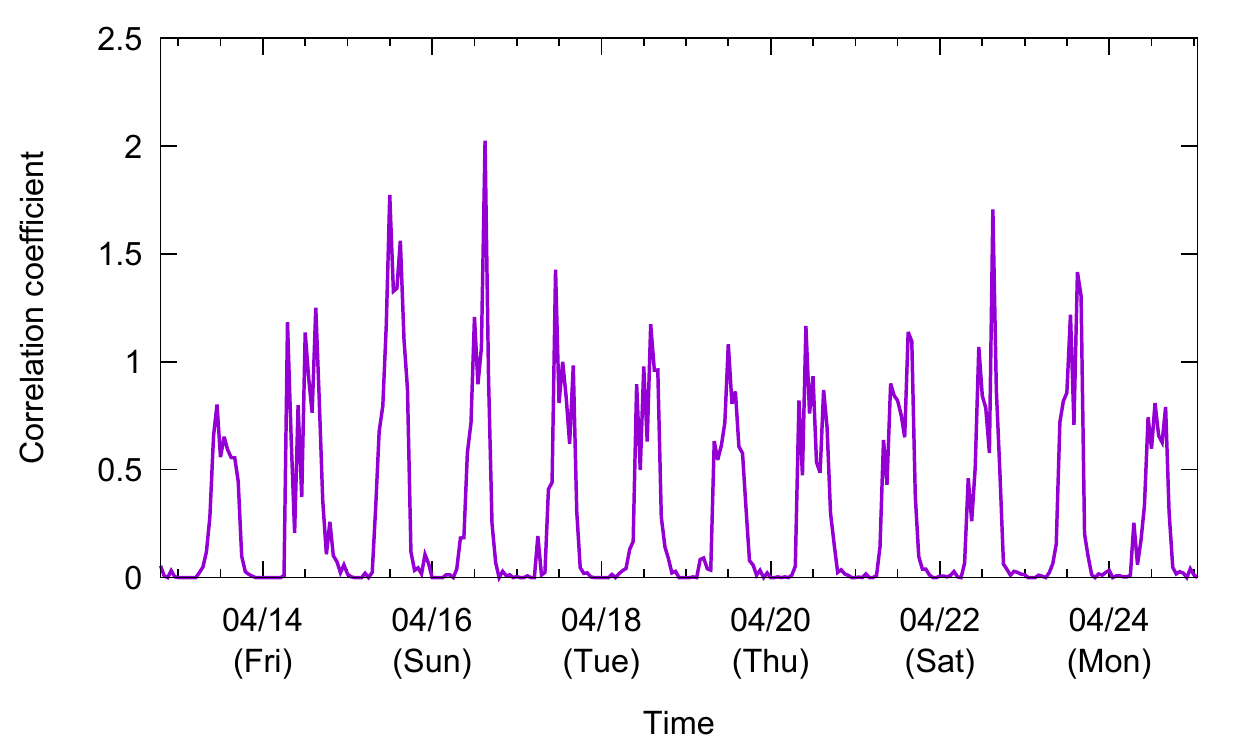} \\
(a) The deployment in Re:START mall in Christchurch. & (b) The correlation coefficients ($a_i$). \\
\includegraphics[width=0.45\linewidth]{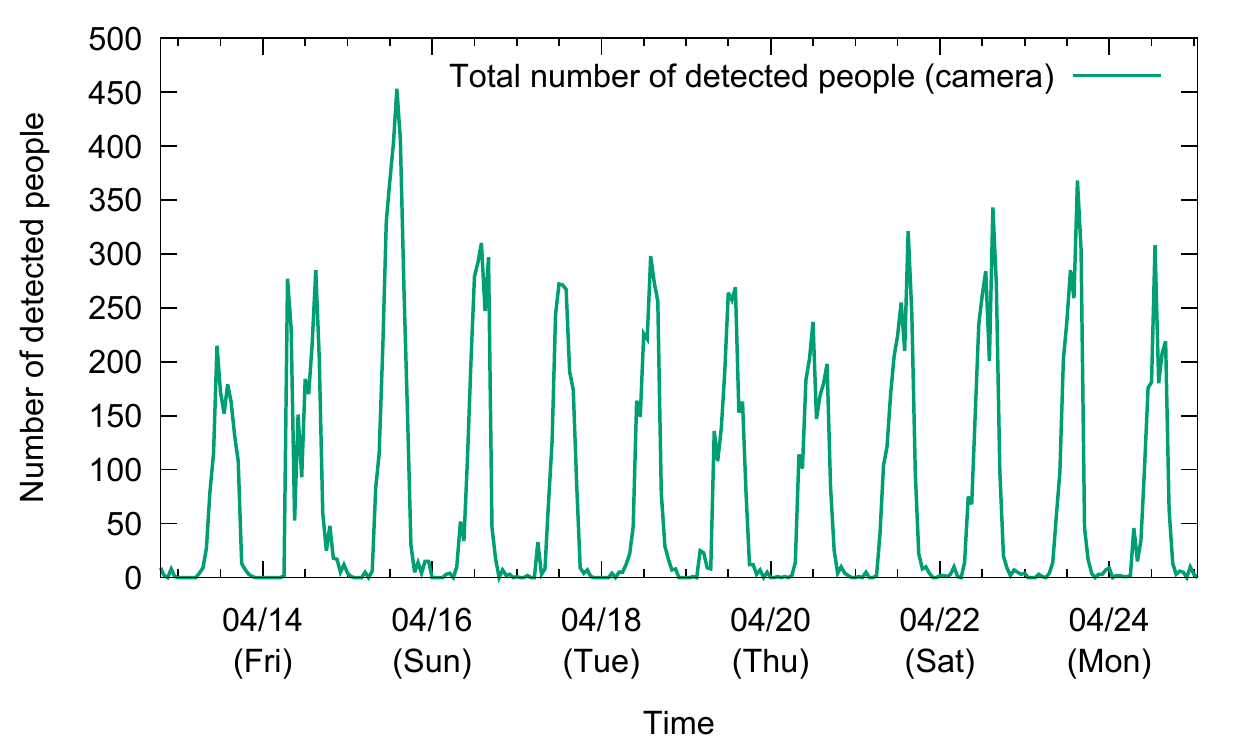} &
\includegraphics[width=0.45\linewidth]{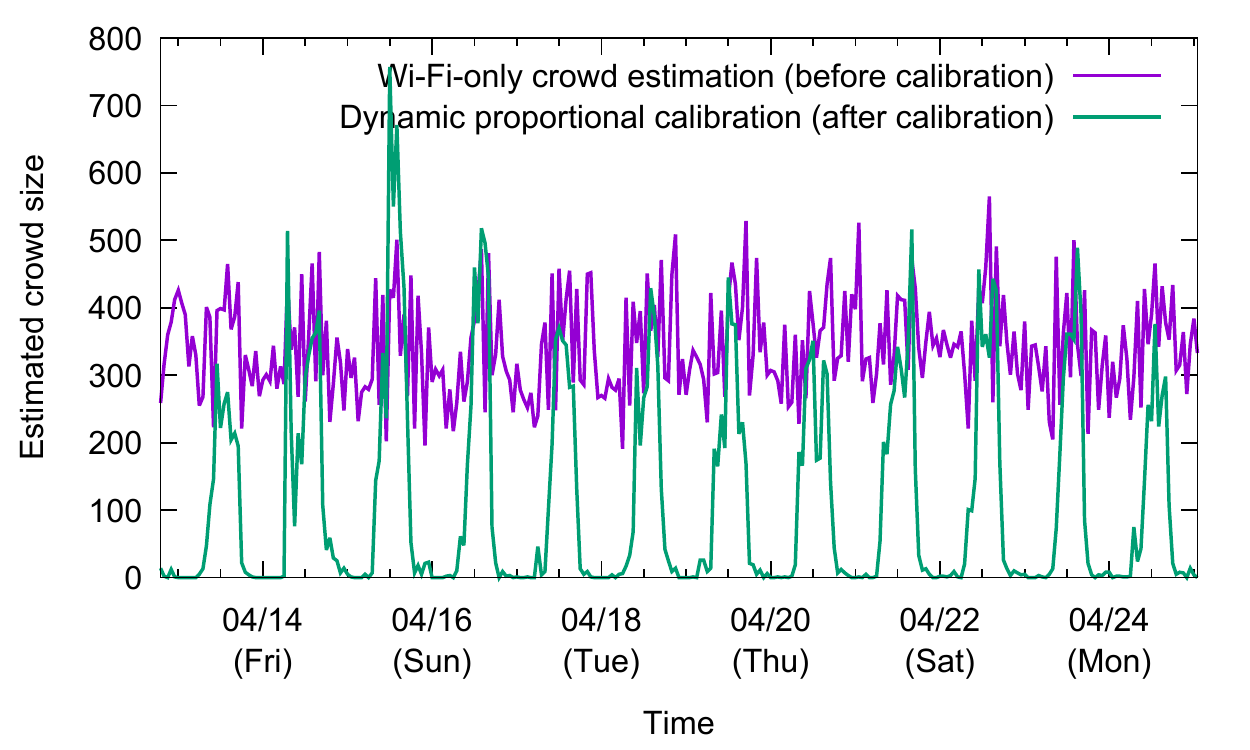} \\
(c) The near ground-truth using the stereoscopic camera. & (d) The calibration results of the Wi-Fi sniffer 4. \end{tabular}
\caption{The deployment and experimental results in the Re:START mall.} \label{Fig:CHCH-deployment}
\end{figure*}

\subsubsection{The IoT Platform}
The crowd estimation service is integrated with the in-house IoT platform. The IoT platform is FIWARE-based and, it includes components that are defined as {\em generic enablers} (GEs) in the FIWARE ecosystem~\cite{FIWARE}. In particular, the components implement \emph{IoT Broker} and \emph{IoT Discovery} GEs of FIWARE. Mainly, IoT Broker is used for distribution of the information coming from IoT data providers (e.g., devices) to the IoT data consumers (e.g., applications). IoT Discovery is necessary for discovering the availability of resources (context). We develop the defined features of the GEs in our open-source IoT components. This paper uses the lightweight \emph{thin broker} and \emph{thin discovery}. They use the same NGSI-9 and NGSI-10 interfaces for real-time services. The main functions of the thin broker are listed below.\\
$\bullet$ \emph{Context query}: Access context information (for data consumers).\\
$\bullet$ \emph{Context subscription}: Subscribe for a change or an update of a context (for data consumers).\\
$\bullet$ \emph{Context notification}: Send notifications to the subscribers (from IoT Broker to data consumers).\\
$\bullet$ \emph{Context update}: Send new context (from data providers to IoT Broker).

Context query returns the latest available data to the data consumer. In the crowd estimation service, it basically returns the latest estimated crowd size. Context subscription is saved in the thin broker whenever a change in the context happens (e.g., new estimated crowd size), a context notification is triggered and the subscriber is notified with an HTTP post. Context subscription includes a reference URL which defines an HTTP server to listen upcoming notifications from the thin broker. Context notification is the result of a context subscription, such that when a change in the subscribed context happens, a notification is automatically triggered. Context update is from the data providers to the thin broker. In the crowd estimation service, the new results for different entities (i.e., Wi-Fi sniffers) are pushed to thin broker every time window.

\begin{figure*}[h!]
\centering
\includegraphics[width=0.8\linewidth]{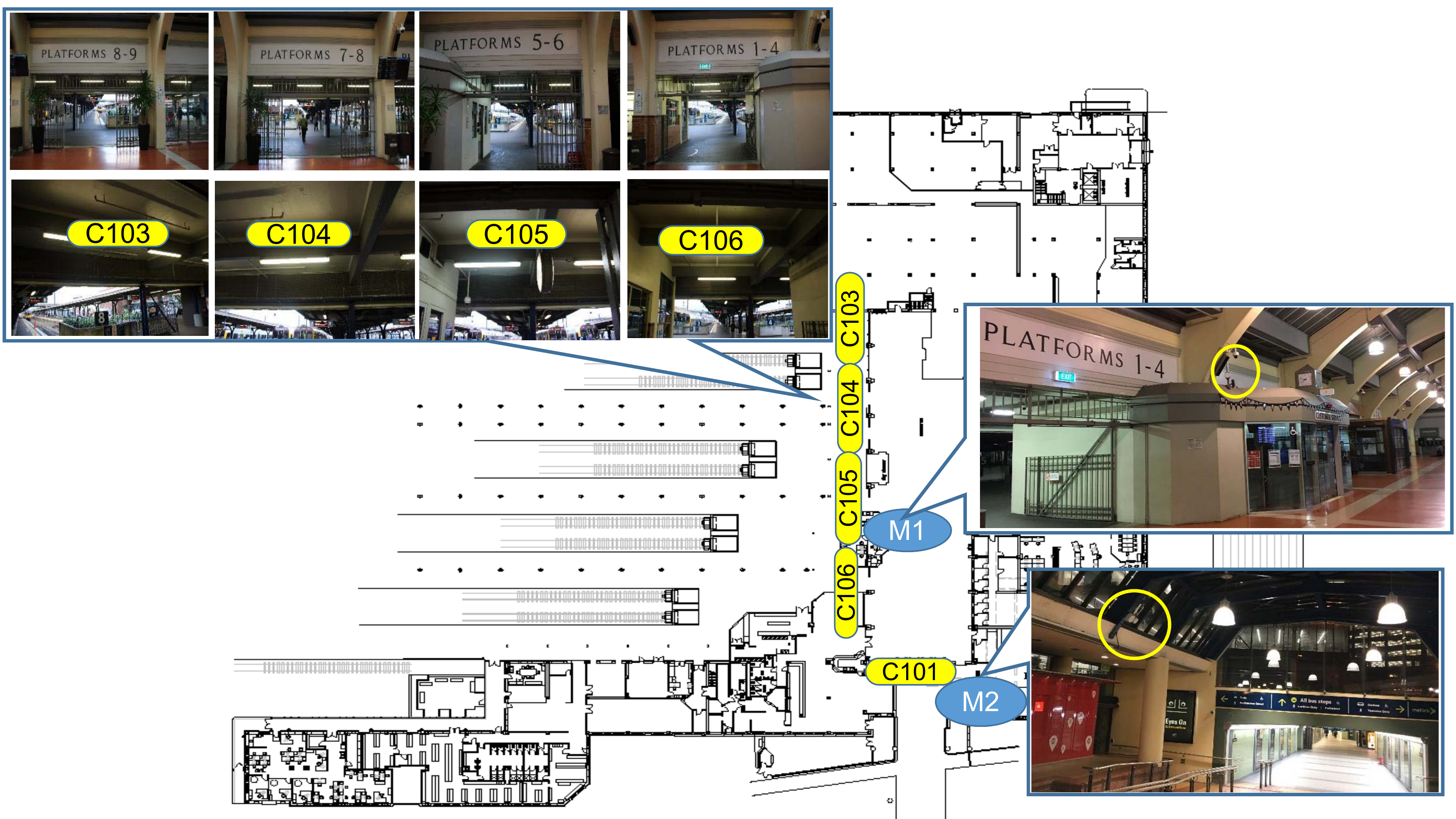}
\caption{Deployment in the Wellington Railway Station.} \label{Fig:WCC-Pilot}
\end{figure*}

\subsection{An Outdoor Pilot Study: Pedestrian Areas}
The first pilot study is conducted in the Re:START mall which is a pedestrian shopping mall in Christchurch, New Zealand from 12-25 April 2017. Before the April, pre-pilots are conducted from December 2016 to April 2017 for device installation, real-time communication testing, and making observations on collected data. The lessons from the pre-pilots are discussed in \Sec{Sec:discussion}. Finally, five Wi-Fi sniffers are deployed in the Re:START mall. \Fig{Fig:CHCH-deployment}(a) shows the deployment, where the shopping mall is built by many containers which divide the entire area into several pedestrian walking areas. The Wi-Fi sniffers are mounted on these containers. The Wi-Fi sniffer 2 and the stereoscopic camera are deployed at the calibration choke point which is located at the main street in the pedestrian shopping mall. As we can see in \Fig{Fig:CHCH-deployment}(b), there exists a regularity in the daily patterns of correlation coefficients between the Wi-Fi-only crowd estimation results and the camera-based people counting results. \Fig{Fig:CHCH-deployment}(c) shows the near ground-truth using the stereoscopic camera in the main street. \Fig{Fig:CHCH-deployment}(d) shows the crowd estimation results before and after the dynamic proportional calibration is applied to the sensing zone covered by the Wi-Fi sniffer 4. The Wi-Fi-only crowd estimation before calibration is mostly overestimated, and the number of detected mobile devices even at midnight is still very high. After applying the proposed algorithm, the calibration results indicate similar daily mobility patterns of the near ground-truth.

\subsection{An Indoor Pilot Study: A Train Station}

\begin{table}
\footnotesize
\caption{The accuracy of stereoscopic cameras.}
\centering
\begin{tabular}{|c|c|c|c|c|}
  \hline			
  IDs      & C103 & C104 & C105 & C106 \\
  \hline
  Accuracy & $85\%$ & $85\%$ & $93\%$ & $95\%$ \\
  \hline
\end{tabular}
\label{table:camera-accuracy}
\end{table}

\begin{figure*}[t!]
\centering
\begin{tabular}{cc}
\includegraphics[width=0.45\linewidth]{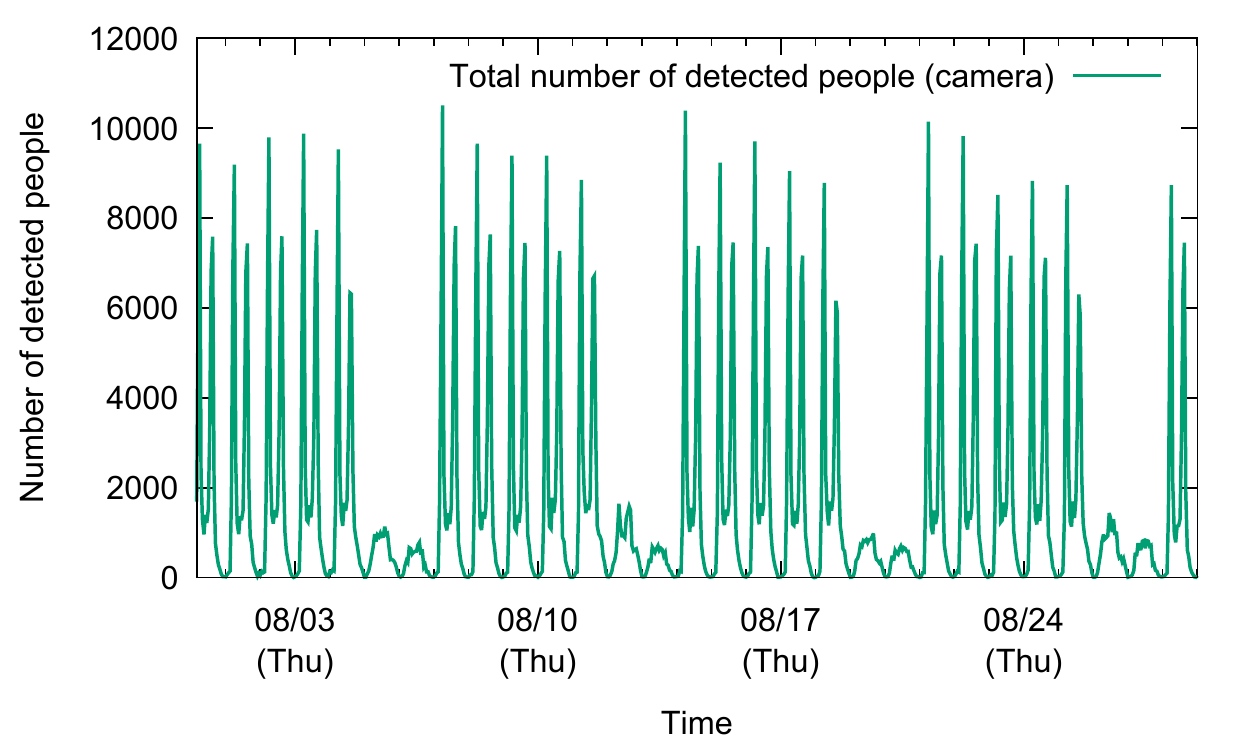} &
\includegraphics[width=0.45\linewidth]{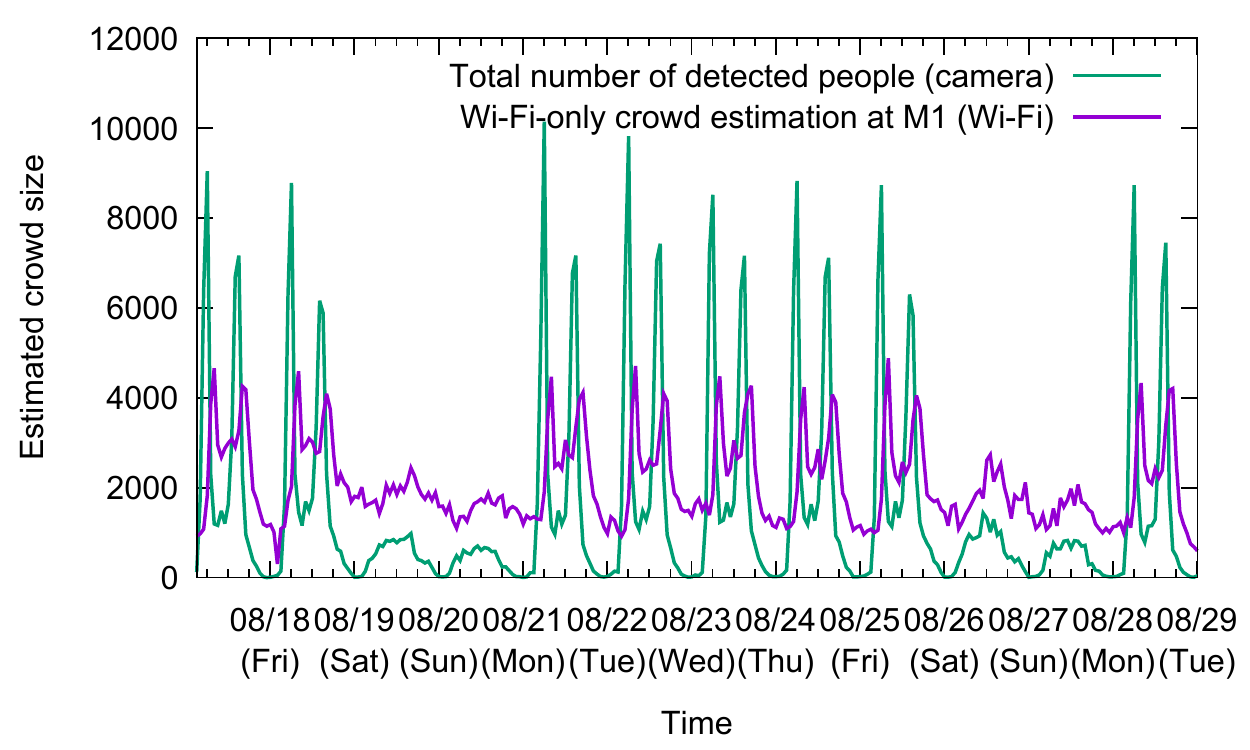} \\
(a) The weekly pattern of the near ground-truth.& (b) The observed daily pattern at the calibration choke point.\\
\includegraphics[width=0.45\linewidth]{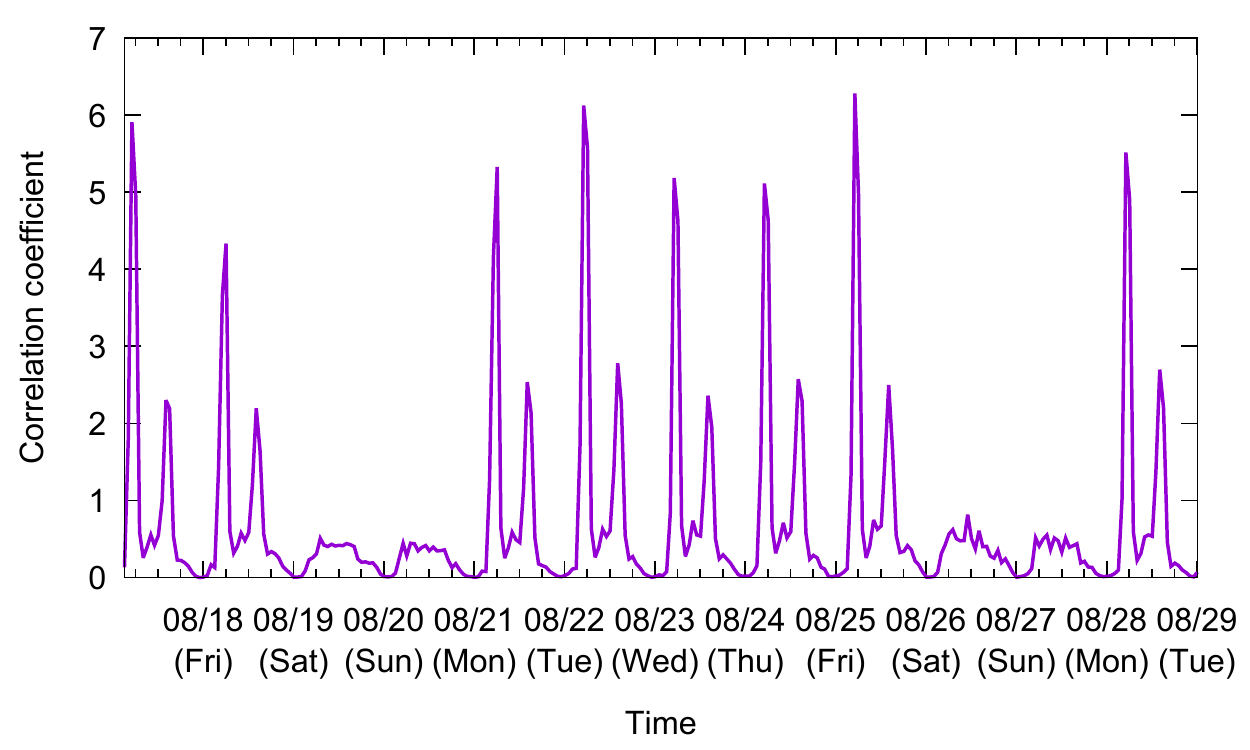} &
\includegraphics[width=0.45\linewidth]{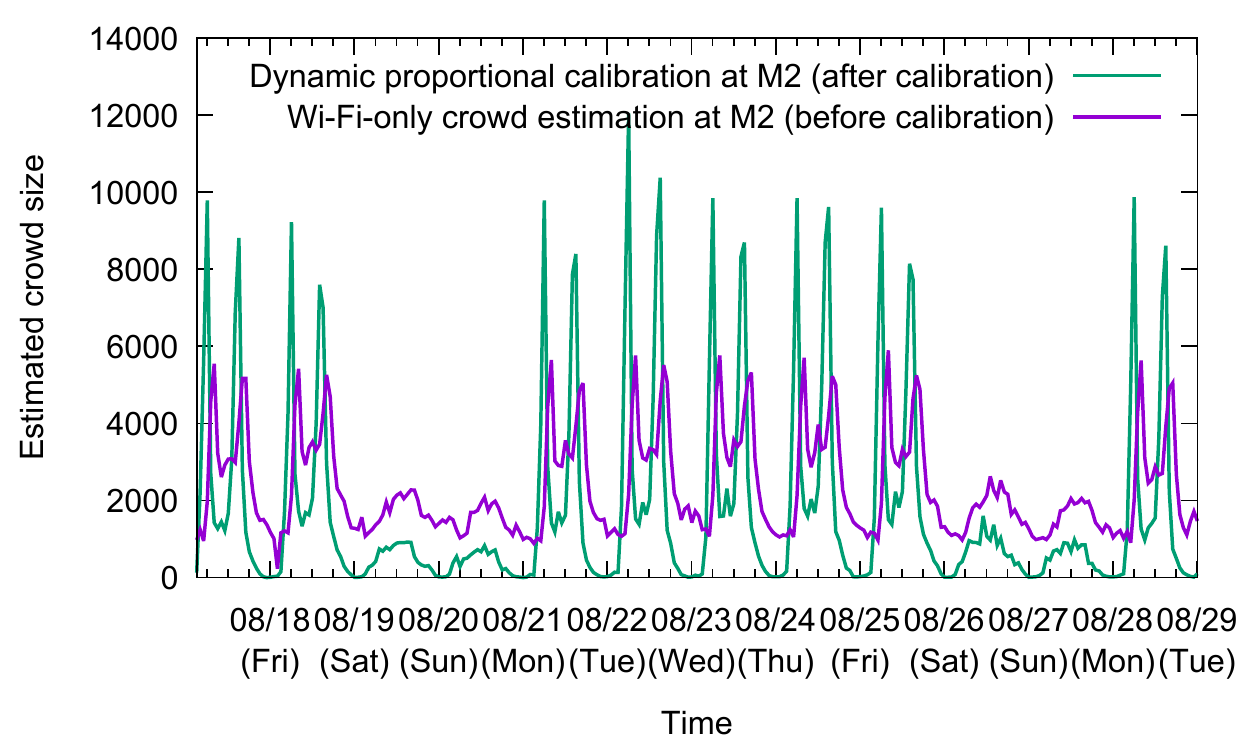} \\
(c) The correlation coefficients at the calibration choke point.& (d) The calibration results of M2.\\
\end{tabular}
\caption{The experimental results in the Wellington Railway Station. 
}
\label{Fig:WCC-ExpResults}
\end{figure*}

An indoor pilot study is conducted in the Wellington Railway Station, New Zealand during 03-24 August 2017. Compared to pedestrians in a shopping mall, the passengers in the train station generally make fast movements. \Fig{Fig:WCC-Pilot} shows the deployment in the train station with multiple entrances. The entrances are considered to form a single calibration choke point, where the passengers moving to/from the platforms are monitored. Two Wi-Fi sniffers M1 and M2 are deployed to capture Wi-Fi probes of mobile devices carried by passengers. The Wi-Fi sniffer M1 is deployed at the calibration choke point, and the Wi-Fi sniffer M2 is located at the side entrance of the train station in the canopy/subway area for monitoring people walking outside the train station. Four stereoscopic cameras are deployed at the calibration choke point for collecting the near ground-truth. Since the entire platform areas consist of multiple platforms, where multiple of them share an entrance, four stereoscopic cameras C103, C104, C105, and C106 are grouped into a ``virtual'' one to cover all entrances of the entire platform areas.

In the indoor scenario, videos from all the stereoscopic cameras are recorded. We manually count the actual number of people in the recorded videos to verify the accuracy of these stereoscopic cameras in the pilot study. \Tab{table:camera-accuracy} indicates the accuracy of the stereoscopic cameras at the calibration choke point. They can provide a minimum accuracy of $85\%$ in the indoor environment.

\begin{figure}
\centering
\includegraphics[width=\columnwidth]{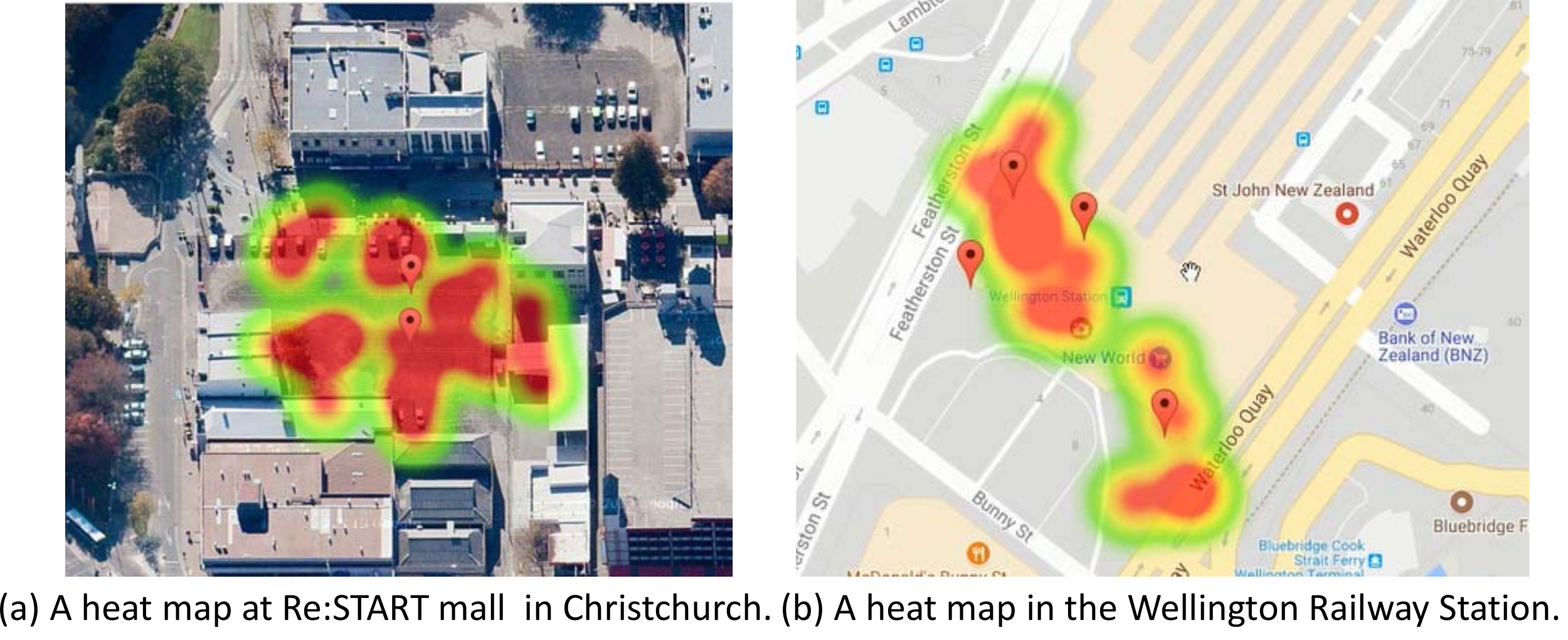}
\caption{Visualization in the two pilot studies.} \label{Fig:UI}
\end{figure}

\Fig{Fig:WCC-ExpResults}(a) shows the weekly pattern of the near ground-truth collected at the calibration choke point. Since the four stereoscopic cameras are grouped into a virtual one, the total number of passengers detected by all of stereoscopic cameras are accumulated for the near ground-truth. As we can see, the numbers of passengers passing though the platform areas during weekdays are much higher than the weekends. \Fig{Fig:WCC-ExpResults}(b) shows the daily pattern of the Wi-Fi-only crowd estimation and the near ground-truth detected by the stereoscopic cameras at the calibration choke point. There exist a peak during commuting time every morning and a sub-peak during the commuting time every afternoon. As it can be seen, the Wi-Fi-only approach underestimates crowd sizes during peak hours, whereas it overestimates crowd sizes during non-peak hours and weekends. The environmental conditions are more dynamically changing compared to the outdoor pedestrian shopping mall in Christchurch. \Fig{Fig:WCC-ExpResults}(c) shows the correlation coefficients between the Wi-Fi-only crowd estimation results and people counting by these stereoscopic cameras at the calibration choke point. It is similar to the weekly mobility pattern in the train station. \Fig{Fig:WCC-ExpResults}(d) shows the calibration results after the dynamic proportional calibration is applied. Since the results are adaptively calibrated based on the near ground-truth, the underestimation and overestimation situations can be mitigated after calibration. \Fig{Fig:UI} shows the heat map views which are included in the visualization dashboards for the two pilot studies.

\subsection{Advanced Performance Comparison}
\begin{figure}
\centering
\includegraphics[width=\columnwidth]{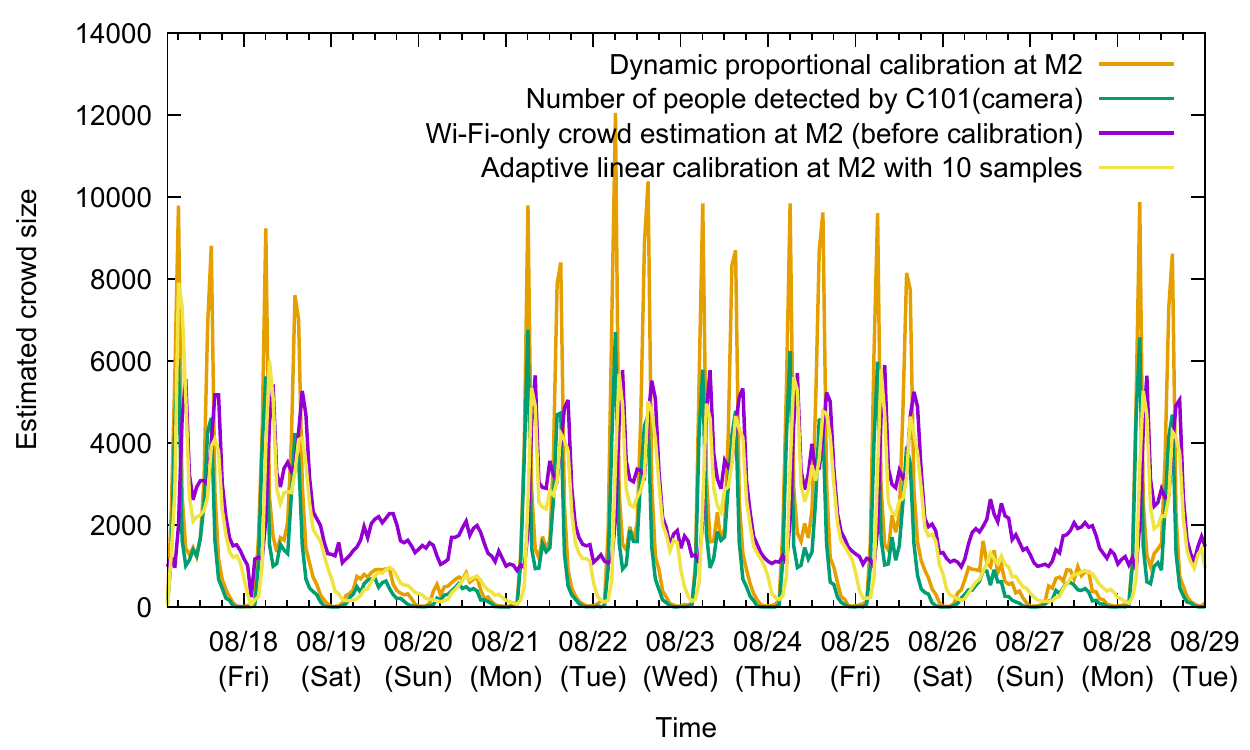}
\caption{Comparison between different calibration algorithms.} \label{Fig:ComparisonDifferentAlgo}
\end{figure}

To verify the accuracy the proposed calibration algorithms, an additional stereoscopic camera C101 is installed next to the Wi-Fi sniffer M2 in the Wellington Railway Station, as shown in \Fig{Fig:WCC-Pilot}. The C101 provides the near ground-truth in the M2's sensing zone to compare with calibration results of the proposed calibration algorithms and verify their accuracy. \Fig{Fig:ComparisonDifferentAlgo} shows the calibration results when the two proposed calibration algorithms are applied to the M2's sensing zone. The dynamic proportional calibration results are closer to the near ground-truth than the adaptive linear calibration results most of the time. It relieves the overestimation situations compared to the Wi-Fi-only crowd estimation. However, dynamic proportional calibration is too sensitive to the extreme changes of correlations during peak hours. By contrast, the adaptive linear calibration can provide more accurate results during peak hours.

\begin{figure}
\centering
\includegraphics[width=\columnwidth]{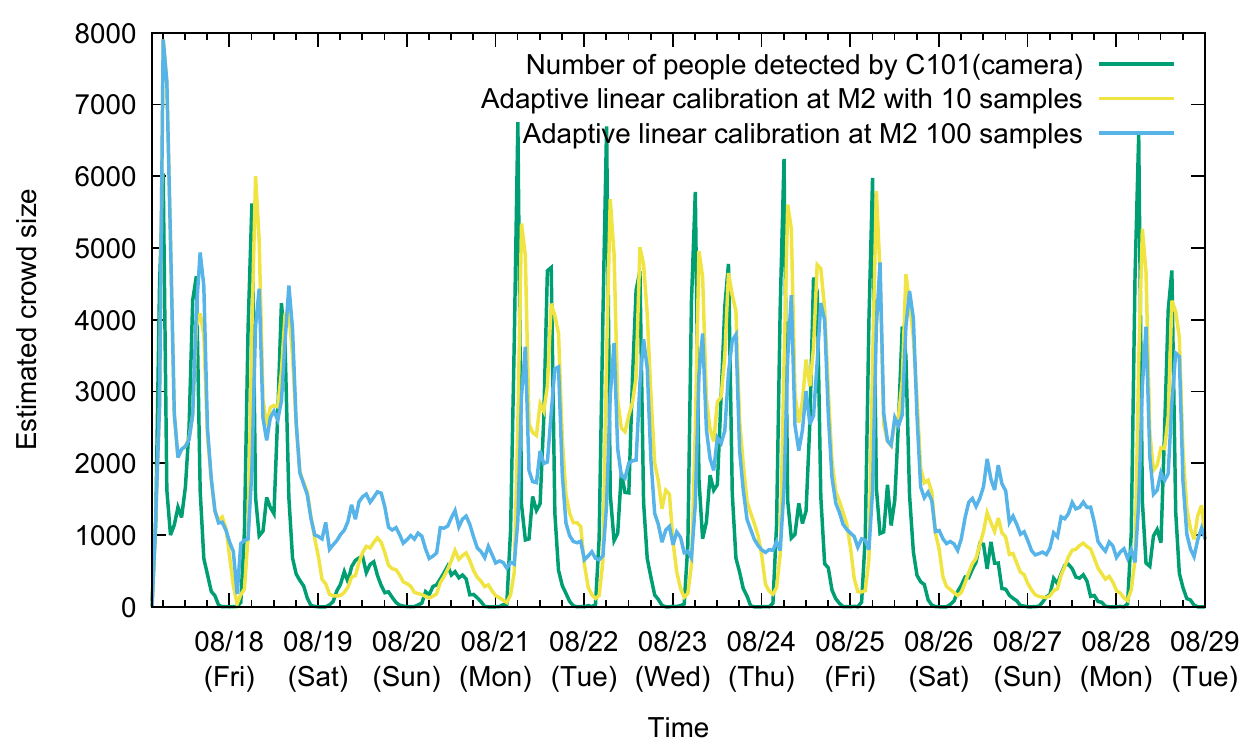}
\caption{Adaptive linear calibration results with different numbers of training data points.} \label{Fig:Comparison-with-different-q}
\end{figure}

Next, we investigate how the number of training data points (i.e., the value of $q$) affects the results when the adaptive liner calibration is adopted. The value of $q$ is changed to 10 and 100 respectively in the experiments. \Fig{Fig:Comparison-with-different-q} shows the experimental results. Interestingly, having more training data points is not always good or necessary especially for an environment with more uncertainties. A larger value of $q$ is not flexible to adapt to extreme changes of correlations during peek hours and midnights because the linear functions at different time are almost fixed. However, having a smaller value of $q$ opens more flexible opportunities for updating the linear functions. This makes the linear functions fit better to the real-time changes in the environment. Therefore, a smaller value of $q$ offers the capabilities to adapt to the dynamic changes in the real-time system.

\begin{figure}
\centering
\includegraphics[width=\columnwidth]{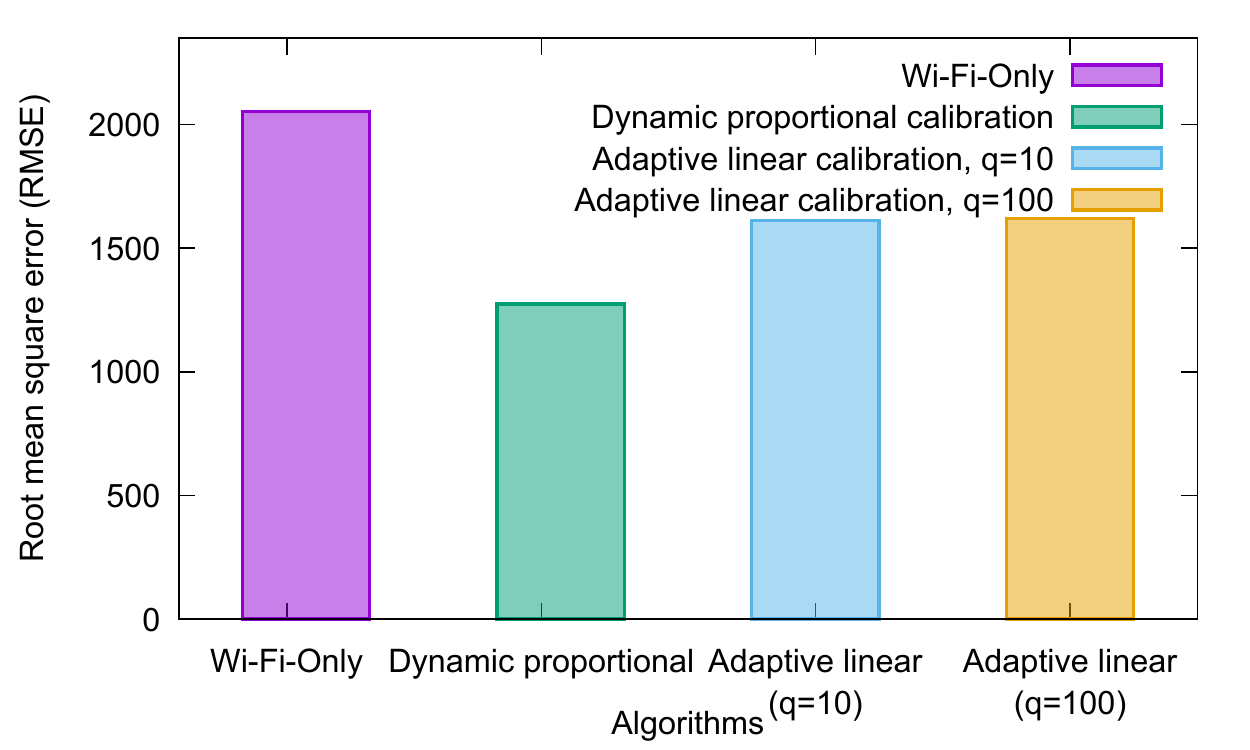}
\caption{RMSEs when different algorithms are applied.} \label{Fig:RMSE}
\end{figure}

\begin{figure}
\centering
\includegraphics[width=\columnwidth]{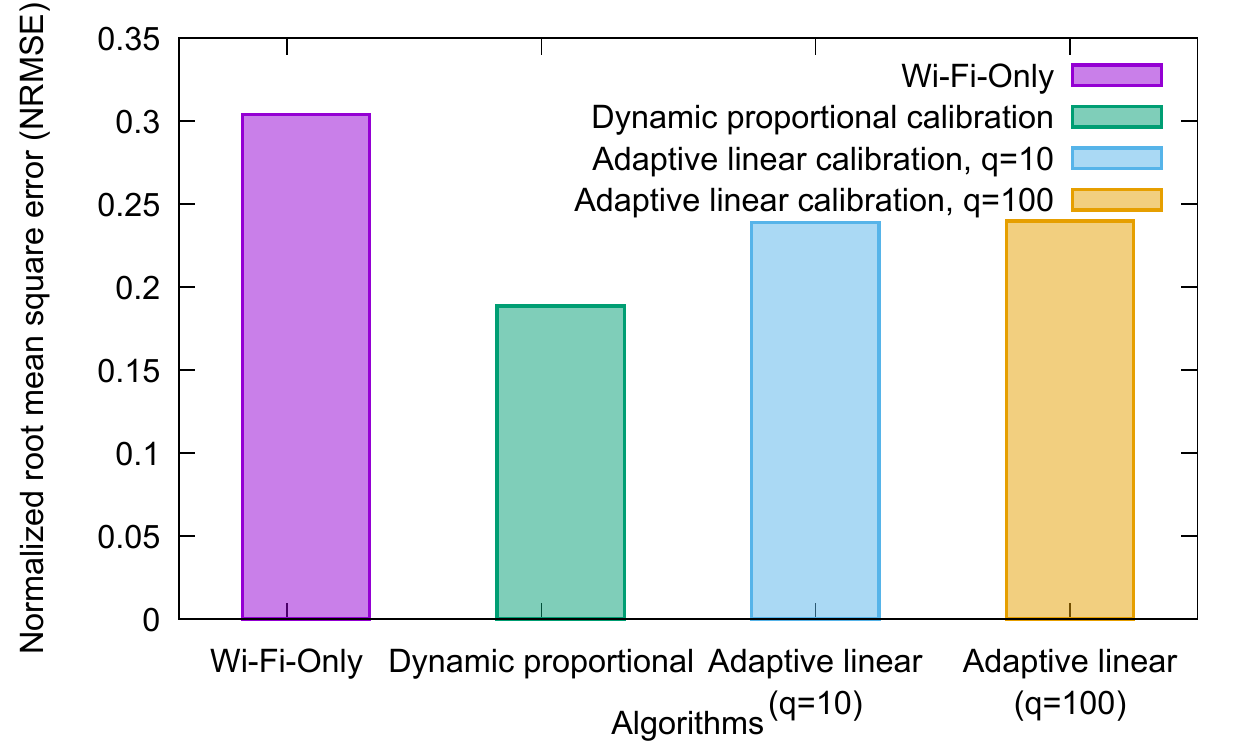}
\caption{NRMSEs when different algorithms are applied.} \label{Fig:NRMSE}
\end{figure}

Then, to quantify the accuracy, the root mean square errors (RMSEs) and the normalized root mean square errors (NRMSEs) are calculated when different algorithms are applied to the collected dataset. For each algorithm, we calculate $RMSE=\sqrt{\frac{\sum_{i=1}^{s} (\tilde{e_{i}}-g_i)^2}{s}}$, where $s$ is the total number of the time windows in the dataset, $\tilde{e_{i}}$ is the crowd estimation using a particular algorithm at the time window $t_i$, and $g_i$ is the near ground-truth provided by the stereoscopic camera C101 at the time window $t_i$. Then, the NRMSE is defined by $NRMSE=\frac{RMSE}{\max_{i=1}^{s}{g_i}-\min_{i=1}^{s}{g_i}}$.
\Fig{Fig:RMSE} and \Fig{Fig:NRMSE} show the evaluation results of RMSEs and NRMSEs. Both of the two proposed calibration algorithms improve the accuracy of crowd estimation compared to the Wi-Fi-only approach. The proposed calibration algorithms can reach a maximum normalized root mean square error of 0.25. Overall, the dynamic proportional calibration provides better crowd estimation accuracy compared to the other algorithms. It also incurs lower computational complexity compared to the adaptive linear calibration which requires a historical set of training data points. \Table{table:Statistics} shows the statistics of errors compared to the near ground-truth when different approaches are applied. The proposed calibration algorithms reduce an average error of $43.68\%$ compared to the Wi-Fi-only approach.

\begin{table*}
\small
\caption{The statistics of errors.}
\centering
\begin{tabular}{|c|c|c|c|c|c|c|c|}
  \hline			
  Algorithms      & Mean & Standard deviation & Minimum & The 1st quartile& The 2nd quartile& The 3rd quartile & Maximum\\
  \hline
  Wi-Fi-Only            & 1303	&1589	&-4919	&1040	&1373	&1825	&4875\\
  \hline 	
  Dynamic proportional	& 659	&1093	&-28	&69	    &245	&590	&5699\\
  \hline
  Adaptive linear(q=10)	& 686	&1462	&-4190	&84	    &427	&1340	&5694\\
  \hline
  Adaptive linear(q=100)& 734	&1447	&-5546	&635	&880	&1181	&5694\\
  \hline
\end{tabular}
\label{table:Statistics}
\end{table*}

\section{Discussion}\label{Sec:discussion}
We discuss technical limitations, deployment issues, experience from the real-world pilots, and future work.

\textbf{Limitations of single-modal technology}: This work is motivated by the findings from our earlier pre-pilots using single-modal technology which motivate us to design the multi-modal approach. With the proposed approach, the two types of sensing technologies can compensate each other's essential limitations. The Wi-Fi-only technology has unstable and invisible coverage due to nature of wireless signals, whereas vision-based technology offers visible coverage which makes verification with the real ground-truth possible. The wireless signals and packets are not reproducible even though the events of crowd appearance and environmental conditions are the same. By contrast, with the same events, measurements of people counting are repeatable and reproducible with vision-based technology. Wi-Fi-only approach may result in a lower accuracy due to uncertain mobility speeds, multiple devices carried by a single person, and more complex environmental conditions, whereas the vision-based technology offers deterministic results. On the other hand, the Wi-Fi-only technology offers a better solution to privacy preservation compared to the vision-based technology and might be more applicable to many regions due to legal regulations. The Wi-Fi-only technology is more flexible for large-scale use-cases required by various stakeholders, whereas the vision-based technology may be limited by the brightness, sizes of targeted areas, and invisibility due to nearby obstacles. The Wi-Fi-only technology offers low-cost deployment and incurs the less communication overhead for data collection compared to collecting videos or images.

\textbf{Requirements for deployment}: The pilot studies are conducted in pedestrians areas without much vehicle traffic (such as a train station with a few entrances and a shopping mall with a main entrance area). The targeted environments have some burst pedestrian traffic such as sport events and daily commuters. A main junction or entrance is preferred to be selected as the calibration choke point due to having more chances to capture most of pedestrians. Distances between the calibration choke point and other Wi-Fi sniffers are not far, and the distributions of pedestrians in these zones are similar. Thus, the correlation learned from the calibration choke point is applicable to these zones. Our system uses the off-the-shelf stereoscopic cameras for counting people \cite{stereoscopic-camera}. To collect the near ground-truth, the stereoscopic cameras should look vertically downwards and visually cover all possible passage areas. This can be achieved by mounting the cameras on the ceilings below which the passage area is not very wide. These constraints limit the usage of stereoscopic cameras to only certain choke points where people are supposed to pass through such as the entrance gates of a shopping mall. However, the proposed algorithms are not limited to the stereoscopic cameras. Other off-the-shelf cameraes can be used to perform people counting \cite{MobotixCameras}\cite{PanasonicCameras}. Alternative options for vision-based people counting could exploit some of the existing real-time object detection approaches \cite{YOLO_objectDetection} with normal CCTV cameras. Privacy-by-design mechanisms are implemented in the proposed system at a low cost, where the privacy-sensitive data in Wi-Fi packets is pre-processed as anonymous data by hashing and salting mechanisms before the proposed algorithms are applied.

\textbf{System limitations and lessons from real-world pilots}: This paper proposes dynamically applying the learned correlations at the calibration choke points to larger scales with less costly Wi-Fi sniffer deployment. While a correlation at a calibration choke point may not be fully applicable to all neighboring zones, it still provides higher accuracy as shown in the experimental evaluation. Our pilot experience shows us that correlating Wi-Fi sniffers with cameras has certain limitations. For instance, in a crowded open area which does not have certain entrance gates such as a beach area, the stereoscopic camera's results cannot be regarded as the near-ground truth. To make verification possible, camera deployment should cover most entrance areas for capturing the near ground-truth at the calibration choke points. For example, a ``virtual'' camera is formed by multiple cameras to cover all entrances of the platforms in the Wellington Railway Station. As expected, areas having many entrance points cause higher costs of deployment. Although the proposed approach is applicable to various medium to large-scale urban areas, it does not take the places with heavy vehicle traffic into account. Considering more complex environments with vehicles could be a future research direction.

\section{Conclusion}\label{Sec:conclusion}
This paper exploits the Wi-Fi probes and computer vision technology to build the crowd estimation IoT service.  This service can provide real-time crowd estimation results across different IoT systems. Using only Wi-Fi probes to estimate crowd sizes may lead to crowd overestimation or crowd underestimation. Auxiliary stereoscopic cameras are introduced to collect the near ground-truth for further calibration. An outdoor pilot study has been launched in the Re:START mall in Christchurch, and an indoor pilot study has been launched in the Wellington Railway Station to verify the developed cross-modal crowd estimation IoT service. The crowd estimation results are available through the IoT platform for supporting diverse real-time IoT applications.
\newpage
\section{Acknowledgment}
\parpic{\includegraphics[width=0.20\linewidth,clip,keepaspectratio]{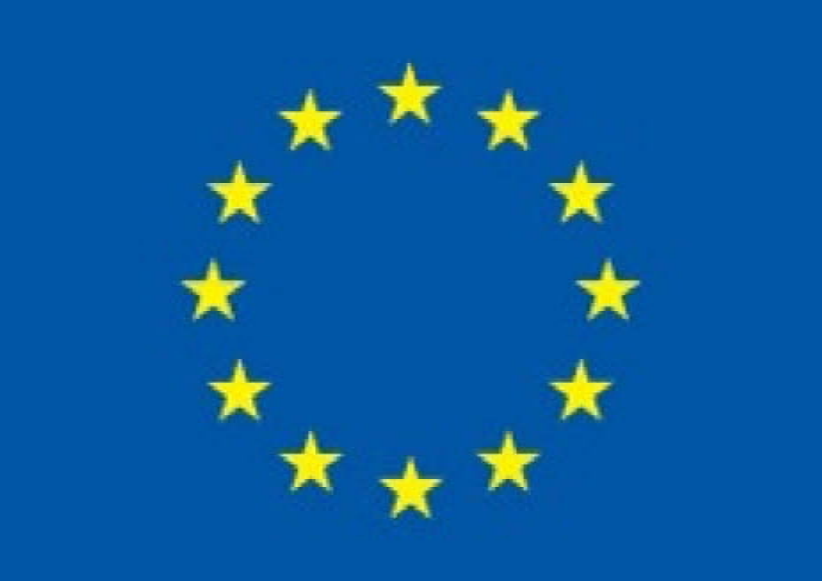}} This work was funded by the joint project collaborations between NEC New Zealand and NEC Laboratories Europe and between NEC Laboratories Europe GmbH and Technische Universit\"at Dortmund, and has been partially funded by the European Union's Horizon 2020 Programme under Grant Agreement No. CNECT-ICT-643943 FIESTA-IoT: Federated Interoperable Semantic IoT Testbeds and Applications. The content of this paper does not reflect the official opinion of the European Union. Responsibility for the information and views expressed therein lies entirely with the authors.

\bibliographystyle{ACM-Reference-Format}                    %

\end{document}